\documentclass[lettersize,journal]{IEEEtran}
\usepackage{amsmath,amsfonts}
\usepackage{algorithmic}
\usepackage{algorithm}
\usepackage{array}
\usepackage[caption=false,font=normalsize,labelfont=sf,textfont=sf]{subfig}
\usepackage{textcomp}
\usepackage{siunitx}
\usepackage{stfloats}
\usepackage{url}
\usepackage{verbatim}
\usepackage{graphicx}
\usepackage{float}
\usepackage{booktabs}    
\usepackage{caption}     
\usepackage{bm}
\usepackage{multirow}
\usepackage{amsmath}
\usepackage{mwe}
\usepackage{wrapfig}
\usepackage{svg}
\usepackage[backend=bibtex]{biblatex}  
\usepackage{diagbox}
\usepackage{tabularx}
\usepackage{comment}
\usepackage{subcaption}
\usepackage{epstopdf}
\usepackage[normalem]{ulem}
\usepackage{bbding}
\usepackage{pifont}
\usepackage{xcolor}

\begin{document}

\title{GDROS: A Geometry-Guided Dense Registration Framework for Optical-SAR Images under Large Geometric Transformations}

\author{Zixuan Sun, Shuaifeng Zhi$^*$, Ruize Li, Jingyuan Xia, Yongxiang Liu, Weidong Jiang

\thanks{All authors are with the College of Electronic Science and Technology, National University of Defense Technology, Changsha, 410073, China.}

\thanks{*Shuaifeng Zhi is the corresponding author: zhishuaifeng@outlook.com.}}

\maketitle

\begin{abstract}
Registration of optical and synthetic aperture radar (SAR) remote sensing images serves as a critical foundation for image fusion and visual navigation tasks. This task is particularly challenging because of their modal discrepancy, primarily manifested as severe nonlinear radiometric differences (NRD), geometric distortions, and noise variations. 
Under large geometric transformations, existing classical template-based and sparse keypoint-based strategies struggle to achieve reliable registration results for optical-SAR image pairs. To address these limitations, we propose GDROS, a geometry-guided dense registration framework leveraging global cross-modal image interactions. First, we extract cross-modal deep features from optical and SAR images through a CNN-Transformer hybrid feature extraction module, upon which a multi-scale 4D correlation volume is constructed and iteratively refined to establish pixel-wise dense correspondences. Subsequently, we implement a least squares regression (LSR) module to geometrically constrain the predicted dense optical flow field. Such geometry guidance mitigates prediction divergence by directly imposing an estimated affine transformation on the final flow predictions. Extensive experiments have been conducted on three representative datasets WHU-Opt-SAR dataset, OS dataset, and UBCv2 dataset with different spatial resolutions, demonstrating robust performance of our proposed method across different imaging resolutions. Qualitative and quantitative results show that GDROS significantly outperforms current state-of-the-art methods in all metrics. Our source code will be released at: \url{https://github.com/Zi-Xuan-Sun/GDROS}.
\end{abstract}


\section{Introduction}
\label{Introduction}
\IEEEPARstart{R}{emote} sensing image registration, which involves aligning images from different sensors, times, and viewing angles, is of utmost importance for improving data representation and enabling seamless multimodal data integration \cite{hughes2020TS-Net,liu2022fast}. Nowadays, with the continuous innovation in sensor technology, remote sensing images have made significant progress in both spatial and temporal resolutions. Among these, optical and SAR sensors, as vital data sources for geospatial information, possess distinct while complementary information characteristics \cite{zhou2024unified, zhang2025multiReview, liweijie2024SARATR}.

Optical imagery captures the shape, color, and texture of surface objects, providing rich visual cues for object recognition and classification. However, as a passive sensing modality, its utility is restricted by solar illumination and is therefore susceptible to data degradation under cloud cover, dense vegetation, or nighttime conditions \cite{liu2024robust, zhang2023optical}. In contrast, SAR imagery, acquired through active radar pulse transmission/reception, captures the backscattering characteristics of targets, revealing subsurface features that are undetectable by optical sensors. The all-weather/day-night operational capability makes SAR sensors particularly suitable for continuous Earth observation. However, SAR image interpretation remains challenging due to complex scattering mechanisms and inherent speckle noise \cite{teng2023omird, sun2024os3flow, liu2024shape}.
The registration of optical and SAR images can effectively enhances geospatial observation capabilities, which has far-reaching implications for tasks such as precision guidance, urban planning, environmental monitoring, and geological surveys \cite{ye2022robust, liu2024softformer, zhang2025m3icnet}.

Numerous studies have been carried out on image registration, including classical algorithms such as SIFT  \cite{ng2003sift}, SURF \cite{bay2008SURF}, and ORB \cite{rublee2011orb}, as well as learning-based methods including SuperPoint \cite{detone2018superpoint}, SuperGlue \cite{sarlin2020superglue} and LoFTR \cite{sun2021loftr}. These approaches predominantly adopt keypoint extraction, description, and matching frameworks, demonstrating satisfactory performance in homogeneous image registration scenarios. However, the substantial modality gap between optical and SAR imagery severely compromises the stability and reliability of keypoint extraction. To address this issue, modality-robust registration algorithms have been explored for optical and SAR images, including classical methods such as OS-PC \cite{xiang2020ospc}, RIFT2 \cite{li2023rift2}, and LNIFT \cite{li2022lnift}, as well as learning-based approaches like MU-Net \cite{ye2022mu-net}, FDNet \cite{xiang2022FDNet}, XoFTR \cite{tuzcuouglu2024xoftr}, and CIRSM-Net \cite{wang2025cirsm}.  Though demonstrating enhanced cross-modal registration accuracy,  these work still persist in employing keypoint matching strategies and are fundamentally constrained by severe geometric transformations. To address this critical limitation, recent research has shifted toward dense correspondence estimation frameworks, predominantly leveraging optical flow techniques such as OSFlowNet-Ft \cite{zhang2023optical} and OS$^3$Flow \cite{sun2024os3flow}. Despite their potential, dense feature-based approaches remain at an early stage of development in optical-SAR image registration (OSIR). Current state-of-the-art methods perform well only within limited range of geometric transformations and have not yet achieved satisfactory performance under large transformations.
Motivated by above challenges, we identify and summarize three critical and persistent challenges inherent to dense feature-based registration strategies in OSIR:

\begin{figure*}[!tb]
\centering
\includegraphics[width=1.0\linewidth, trim=40pt 25pt 30pt 50pt, clip]{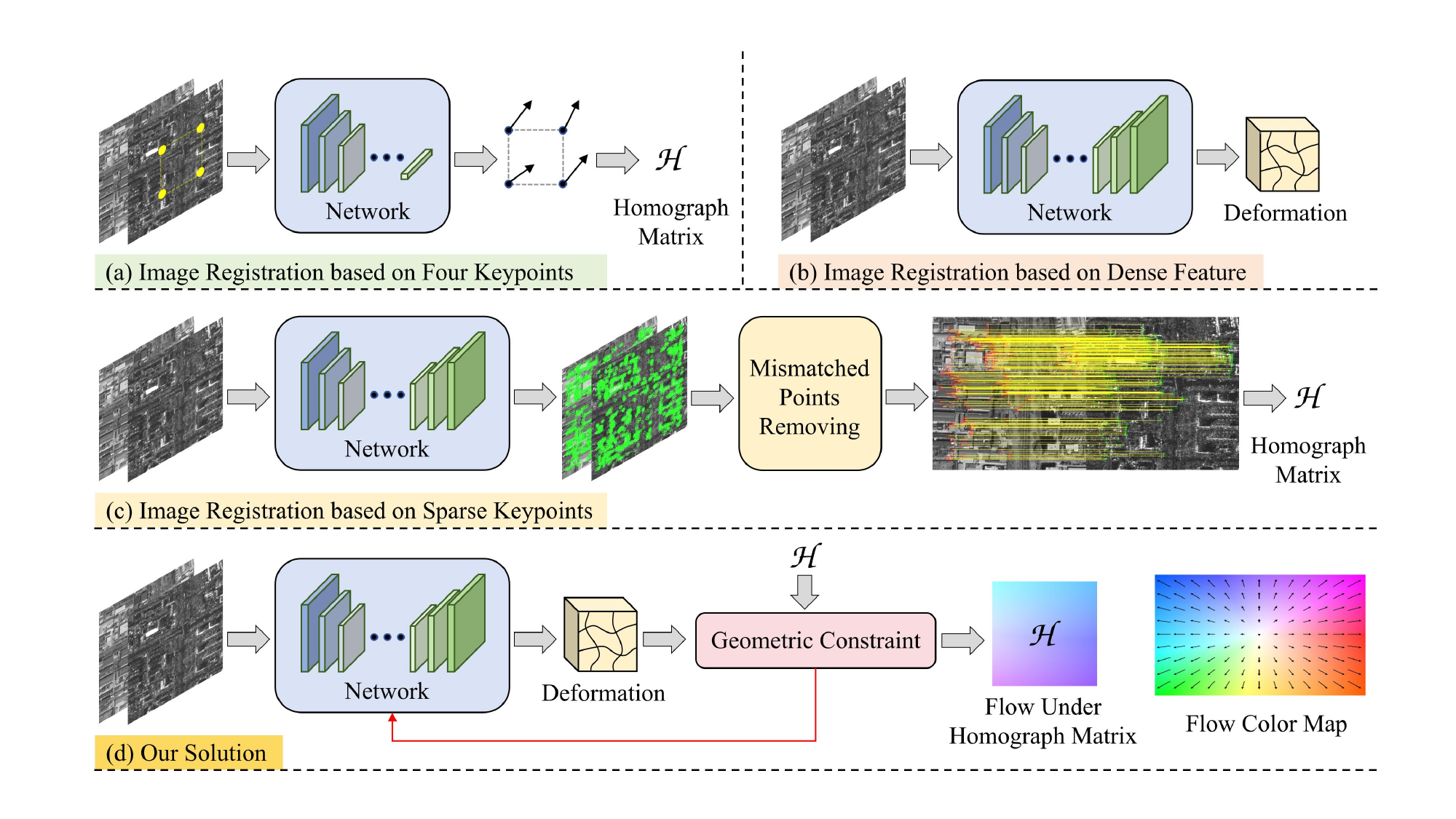}
\caption{Overview of learning-based optical-SAR image registration (OSIR) frameworks.
(a) Predicting motion offsets of four fixed reference points to solve homography/affine matrix, typically employing an encoder-only network architecture.
(b) Describing non-rigid transformations via dense optical flow, typically employing an encoder-decoder network architecture.
(c) Predicting sparse(semi-dense) keypoints correspondences, filtering mismatches, and finally estimating a homography/affine matrix via geometric rectification.
(d) Our proposed solution GDROS: integrating cross-modal dense optical flow with geometric constraints to achieve geometry-guided dense registration.}
\label{fig_overview}
\end{figure*}

\textbf{Challenges in Large Geometric Transformations.}  
Optical-SAR image pairs with significant geometric transformations exhibit large displacement between corresponding points, which intensifies occlusion effects and leads to mismatches, especially for pixels lacking valid correspondences within image boundaries. Furthermore, extensive geometric transformations including large rotations and scaling variations further widens the modality gap between optical and SAR, distorting their spatial structural relationships and similarities. To address above issues, one possible solution is to expand the potential spatial search range, which, however, results in a substantial increase in computational complexity, demanding efficient network architectures.

\textbf{Challenges in Cross-Modal Image Representation.} 
The inherent divergence in optical and SAR imaging mechanisms fundamentally restricts their feature compatibility. For instance, a mountain ridge may exhibit natural undulations in optical imagery but manifest as folded geometries in SAR due to layover effects. Perfect pixel-to-pixel correspondence between optical and SAR images is practically unattainable. 
Dense OSIR approaches predominantly acquire matching similarity using independently extracted image features. However, such strategy fails to capture invariant cross-modal representations or align their feature distributions. Therefore, it remains essential to design effective learning mechanisms for cross-modal representation.

\textbf{Challenges in Filtering Outlier Correspondences.} 
In terms of modeling 6-DoF affine transformations, the redundant degrees of freedom inherent in dense optical flow fields inevitably introduce registration errors. Classical outlier filtering approaches, mostly known as RANSAC \cite{fischler1981ransac}, have greatly improved the robustness of keypoint-based registration, yet are incompatible with dense registration methods. Specifically, the large number of matching candidates and the spatial smoothness within flow fields impose prohibitive computational costs when revealing potential inliers via randomly sampling. Consequently, an effective outliers filtering strategy tailored for dense correspondences remains as an unresolved challenge.

As an attempt to address aforementioned issues, we propose GDROS, an attention-driven dense OSIR framework to predict precise dense correspondences even under challenging large geometric transformations. 
In contrast to existing methods based on attention mechanism, such as XoFTR, GDROS discards the self-attention mechanism and focuses solely on cross-modal interaction to avoid excessive smoothing of intra-modal features. Furthermore, by explicitly modeling affine transformations, GDROS effectively leverages geometric constraints to suppress optical flow mismatch points. Compared to the coarse-to-fine strategies commonly employed in conventional approaches, this explicit affine modeling combined with end-to-end training enables better handling of large geometric deformations, thereby improving the accuracy and robustness of optical-SAR image registration.
Extensive quantitative and qualitative experimental results show that GDROS strikes the leading optical-SAR registration accuracy without introducing much computational overhead. Our main contributions are summarized as follows:

- We propose an end-to-end flow prediction network for dense OSIR by incorporating explicit prior of affine transformation. Most notably, our method demonstrates superior performance against leading baselines, especially on optical-SAR images with large geometric transformations.

- We introduce a cross-modal feature fusion module via a dual-level mutual attention mechanism. This design effectively bridges the domain gap between optical and SAR modalities, enhancing their inter-modal similarity and improving alignment precision.

- The proposed differentiable regression module employs a least-squares formulation by estimating 6-DoF affine transformation parameters for refined dense flow fields. It implicitly removes outliers deviating affine transformations and thus significantly enhances registration reliability.

The rest of this paper is organized as follows. Section~\ref{Related Work} introduces related studies on OSIR. Section~\ref{Approach} introduces the proposed GDROS. Section~\ref{Experiments} conducts extensive experiments and analysis on the effectiveness of our architecture and its robustness to image resolution.Section~\ref{Conclusion} concludes the paper and discusses future work.

\section{Related works}
\label{Related Work}
In this section, we first systematically summarize the evolution and inherent limitations of existing sparse-keypoints-based registration methodologies in Section~\ref{sec2.1：keypoints-based registration}. Subsequently, we critically analyze prevailing dense-feature-based registration approaches and their performance bottlenecks in cross-modal scenarios in Section~\ref{sec2.2：Dense-based registration}. Finally, we rigorously outline geometric prior-based outlier filtering strategies for mitigating mismatches under geometric discrepancies in Section~\ref{sec2.3：Prior knowledge}.

\subsection{OSIR based on Sparse Keypoints}
\label{sec2.1：keypoints-based registration}
Sparse keypoint-based registration methods typically involve keypoint extraction, description, matching, and outlier rejection, as illustrated in Fig.~\ref{fig_overview}(a) and (b). Many existing registration methods largely adhere to this paradigm. However, in the task of OSIR, due to modal differences such as geometric distortions, nonlinear radiometric variations, and speckle noise, keypoint detectors often struggle to extract a sufficient number of robust and reliable interest points between images.

To address this challenge, RIFT \cite{li2019rift} employs phase congruency features instead of traditional amplitude- or gradient-based features, enhancing robustness to NRD. RIFT2 \cite{li2023rift2} significantly improves computational efficiency by replacing the Gabor filter module with the Fast Fourier Transform (FFT). LNIFT \cite{li2022lnift} further mitigates the modality gap in optical-SAR pairs through its proposed normalization operation. HOWP \cite{HOWP2023} adopted a feature aggregation strategy to optimize keypoints by separately extracting corner and blob features. MOSS \cite{zhang2024MOSS} leveraged multidimensional oriented self-similarity features to progressively improve registration performance. The SOFT \cite{zhang2025SOFT} method enhanced rotational invariance in matching by constructing a novel second-order tensor orientation descriptor. Nevertheless, these traditional sparse keypoint-based optical-SAR registration methods exhibit inherent limitations, as they fail to fully exploit local texture information through the combination of phase, amplitude, and gradient.

In contrast, learning-based methods demonstrate superior performance in OSIR tasks, leveraging their powerful feature extraction capabilities. TS-Net \cite{hughes2020TS-Net} introduces a three-stage framework for sparse image registration between SAR and optical images. It utilizes deep neural networks (DNNs) to encode region selection, correspondence heatmap generation, and outlier removal. MU-Net \cite{ye2022mu-net} employs a coarse-to-fine registration pipeline by stacking multiple DNN models. It directly computes affine transformation parameters by learning correspondences for four fixed keypoints, as shown in Fig.~\ref{fig_overview}(a). However, while four keypoints suffice for affine parameter calculation, their inherent instability significantly compromises registration accuracy. LoFTR \cite{sun2021loftr} establishes coarse matches between grids and subsequently refines them using fine features, implementing a coarse-to-fine matching strategy. 
Building upon LoFTR, XoFTR \cite{tuzcuouglu2024xoftr} integrates masked image modeling pre-training, fine-tuning, and image enhancement techniques to address the modality gap. However, the matching process in this method heavily relies on feature similarity based on the attention mechanism, rather than explicit geometric constraints. As a result, it is prone to ambiguous matches in areas with repetitive textures, symmetrical structures, or weak textures. Although it relies on RANSAC post-processing to estimate the geometric model, this post-processing step cannot correct the inherent ambiguity in the underlying feature matches.

In summary, sparse keypoint-based optical-SAR registration techniques have achieved considerable progress, characterized by strong feature robustness and high extraction efficiency. However, in large-scale transformation scenarios, the difficulty of extracting transformation invariance features increases substantially, bringing significant challenges to these methods. Dense feature-based optical-SAR registration techniques provide a viable solution to address these challenges. 

\subsection{OSIR based on Dense Feature}
\label{sec2.2：Dense-based registration}
Dense optical flow estimation computes motion displacements for all pixels in an image. By establishing pixel-wise dense correspondences, it circumvents the challenges associated with keypoint extraction, demonstrating significant potential in optical-to-SAR image registration tasks.

Since its inception in the 1950s \cite{gibson1950perception}, the field of optical flow estimation has witnessed substantial progress. Traditional methods, which rely on the brightness constancy assumption, such as the Lucas-Kanade \cite{lucas1981iterative} and DeepFlow \cite{weinzaepfel2013deepflow}, are ill-suited for optical-SAR image pairs due to their significant radiometric and structural differences. Deep learning-based approaches have overcome this limitation through powerful feature extraction capabilities. FlowNet\cite{dosovitskiy2015flownet} served as a pioneering milestone, being the first deep learning model to outperform classical algorithms. Subsequently, network architecture design has emerged as a pivotal factor in enhancing optical flow precision, spurring ongoing research and innovation. PWC-Net \cite{sun2018pwc-net} introduced an enhanced spatial pyramid network that combines traditional stereo matching, feature extraction, and cost volume mechanisms with deep learning methodologies. RAFT \cite{teed2020raft}  innovatively incorporated a Gated Recurrent Unit (GRU) module for iterative updates, mimicking the iterative refinement process of conventional optimization methods, and marked another major milestone in optical flow estimation.

Following these advances, and inspired by the success of Transformers in computer vision, recent studies have begun leveraging the global modeling capacity of Transformers to tackle large-displacement optical flow estimation. Transformer-based models such as GMFlow \cite{xu2023gmflow} focus on global feature similarity by replacing GRU modules with stacked Transformer blocks, achieving performance superior to RAFT. FlowFormer \cite{huang2022flowformer} further improves registration accuracy by utilizing self-attention mechanisms to effectively capture long-range dependencies and spatial relationships among pixels. FlowFormer++ \cite{shi2023flowformer++} proposed a masked cost volume auto-encoding scheme to pre-train the cost volume encoder more efficiently. MemFlow \cite{dong2024memflow} draws on the attention mechanism of Transformers to achieve effective aggregation of historical information, significantly enhancing estimation accuracy and generalization while maintaining real-time performance.

Currently, OSIR methods based on sparse and dense features have achieved significant development, with their continueously increasing performance. However, most of them focus on extracting features from the amplitude and scattering properties of optical and SAR images, lacking geometric constraints from prior knowledge and physical properties, which leads to instability and insufficient robustness in practical scenarios.

\begin{figure*}[!tb]
\centering
\includegraphics[width=1.0\linewidth, trim=175pt 180pt 120pt 80pt, clip]{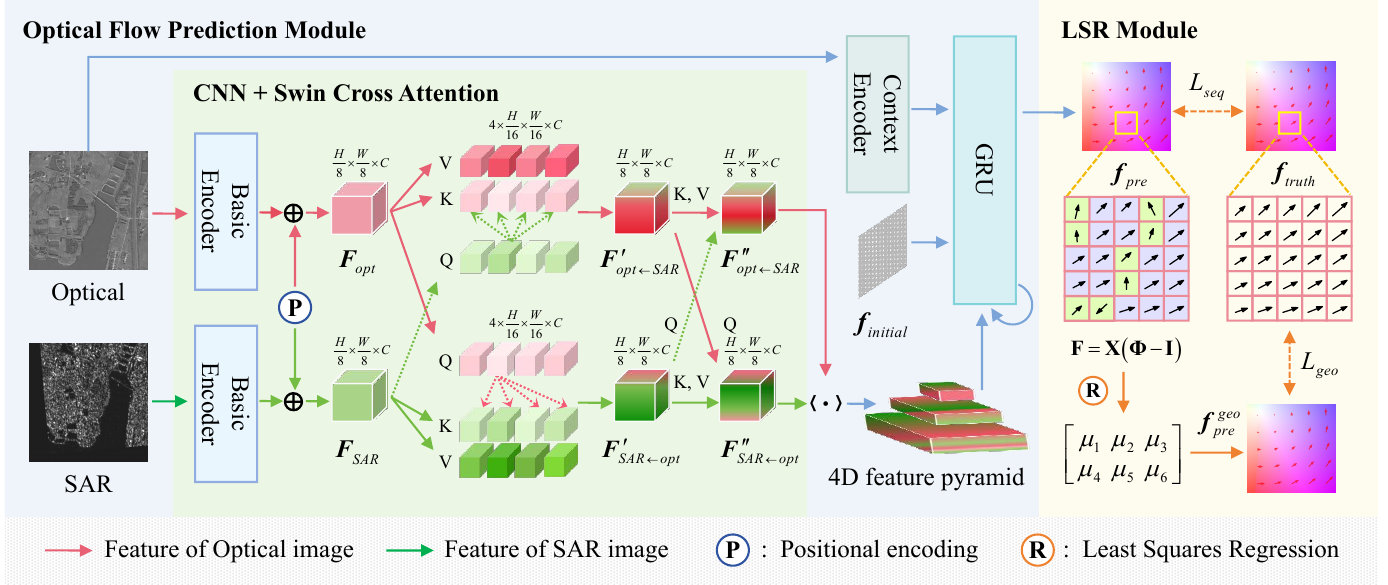}
\caption{Framework of our method GDROS. The input optical-SAR image pairs undergo attention mechanism-enabled feature extraction to obtain two distinct deep feature spaces, ${F}''_{opt\leftarrow SAR}$ and ${F}''_{SAR\leftarrow opt}$, with enhanced inter-modal information interaction, as depicted in the green-highlighted region. By leveraging these deep feature spaces, we construct a multi-scale 4D feature pyramid that enables GRU-based iterative refinement to generate dense optical-to-SAR flow fields. Subsequently, in the LSR-based geometric consistency enforcement module (yellow-highlighted region), geometric consistency constraints are systematically applied to correct mismatches in the initial flow field, ultimately yielding an accurate radiometric transformation model.}
\label{Framework}
\end{figure*}

\subsection{OSIR with Prior Knowledge Constraint}
\label{sec2.3：Prior knowledge}
Outlier filtering, commonly referred to as mismatch removal or correspondence selection, serves to identify geometrically consistent correspondences (inliers) while rejecting spurious matches (outliers) within candidate sets. Within the context of optical-SAR image registration tasks, the fundamental geometric model is typically formulated as a global affine transformation. Under such constraints, outlier rejection predominantly relies on geometric consistency criteria for establishing robust feature correspondences.
The most classical and widely adopted methodology in this domain remains the RANSAC (Random Sample Consensus) \cite{fischler1981ransac} algorithm, which has undergone extensive investigation for decades \cite{derpanis2010ransacoverview}. As an iterative hypothesis-testing framework, RANSAC repeatedly samples minimal subsets of correspondences to hypothesize provisional parameter models, subsequently evaluating model quality through the computation of inlier coherence counts. The algorithm's performance exhibits sensitivity to critical parameters including iteration count and threshold settings. To address these limitations, numerous RANSAC variants have emerged, each introducing innovative sampling strategies and reliability metrics. Notable advancements include MLESAC (Maximum Likelihood Estimation Sample Consensus) \cite{torr2000mlesac} that incorporates probabilistic correspondence weights, PROSAC (Progressive Sample Consensus) \cite{chum2005PROSAC} that utilizes spatial coherence for guided sampling, and MAGSAC (Motion-Aware Generalized Sample Consensus) \cite{barath2019magsac} that integrates motion estimation priors.

These global consensus-based methodologies upon, while theoretically guaranteeing geometric model correctness, are often associated with significant computational overhead that conflicts with real-time image processing requirements. 
Additional alternative frameworks including Neighborhood Consensus Methods, Descriptor-Based Approaches, and Graph-Based Algorithms \cite{shen2024outlier}. However, these technologies exhibit inherent sensitivities to local features and possess fundamental limitations in addressing heterogeneous image registration tasks involving large geometric deformations, and are therefore excluded from further discussion in this paper.

\section{Methodology}
\label{Approach}
In this section, we present the architectural details of our proposed method GDROS. The overall pipeline is shown in Fig.~\ref{Framework} and comprises two primary components: a dense optical flow prediction module and a Least Squares Regression (LSR) module. In Section~\ref{method-sub-A}, we first introduce the dense flow prediction process, followed by a particular emphasis on the proposed cross-modal feature extraction module bridging the domain gap between heterogeneous optical-SAR modalities.
Subsequently, Section~\ref{method-sub-B} presents the LSR module for refined flow predictions under affine transformation constraints, as well as its outlier rejection mechanism.
Finally, Section~\ref{method-sub-C} describes the complete network training strategy.

\subsection{Dence Optical Flow Prediction}
\label{method-sub-A}
Large geometric transformations exacerbate the modality discrepancies between optical-SAR image pairs, making the extraction of reliable and stable keypoints particularly challenging. In contrast, our method opts to predict a per-pixel dense displacement flow field $ f: \mathbb{R}^2 \rightarrow \mathbb{R}^2 $, 
which establishes pixel-wise correspondence between source $ I_s $ and target $ I_t $ images. Specifically, for a pixel located at $ (x_1, y_1) \in \mathbb{R}^2 $ in $ I_s $, the corresponding position $ (x_2, y_2) \in \mathbb{R}^2 $ in $ I_t $ satisfies:
$x_2 = x_1 + f_x(x_1, y_1)$, $y_2 = y_1 + f_y(x_1, y_1)$.

\noindent\textbf{Network Backbone.} Our dense optical flow prediction backbone, as shown in blue-highlighted region of Figure~\ref{Framework}, inherits the RAFT  \cite{teed2020raft} meta-architecture comprising three core components: feature extraction module, 4D volumetric space construction, and a GRU-based iterative refinement module.  It is noteworthy that the context encoder aims to provide essential contextual information for the GRU blocks, and we choose optical images as input given their richer textural and structural details compared to SAR images. 
The GRU update block incorporates the conditioned iterative optimization paradigm from classical approaches, generating a sequence of optical flow $\left\{ f^{1},...,f ^{N}\right\}$ starting from an initial flow field estimation $f_{0}= \textbf{0}$. Through this process, each estimate undergoes progressive refinement via iterative MSE minimization against the ground-truth optical flow field. 

\noindent\textbf{Feature Extraction.} In optical-SAR image registration tasks with significant geometric transformations, a fundamental challenge lies in bridging the modality gap between optical and SAR images while effectively extracting their shared latent structural features. To address this, we propose a hybrid feature extractor (highlighted in green in Fig.~\ref{Framework}) that synergistically integrates convolutional neural networks (CNNs) with Transformer architectures. This configuration preserves intrinsic fine-grained spatial information through CNNs' local receptive fields while enabling long-range cross-modal information exchange via Transformers' attention mechanisms.

Specifically, we employ a weight-sharing ResNet architecture pre-trained on ImageNet as the base encoder to extract domain-specific features $\mathbf{F}_{\mathrm{opt}}$ and $\mathbf{F}_{\mathrm{sar}}$ from optical and SAR images respectively. Consistent with RAFT, we perform 8× spatial downsampling during feature extraction to maintain computational tractability. However, the CNN-derived features operate in isolation, insufficient to overcome the fundamental modality disparity between heterogeneous image domains. To resolve this limitation, we innovatively design a Cross-Attention-Only Transformer module that completely eliminates self-attention operations, structured as follows: 

\noindent\textbf{Positional Encoding.} We embed fixed 2D sinusoidal positional embeddings into CNN-extracted features $\mathbf{F}_{\mathrm{opt}}$ and $\mathbf{F}_{\mathrm{sar}}$, endowing the system with explicit spatial awareness, following standard practice in DETR \cite{carion2020DETR}:
\begin{equation}
\begin{gathered}
\mathbf{F}_{\mathrm{opt}}^{\mathrm{pos}} = \operatorname{Pos}(\mathbf{F}_{\mathrm{opt}}),\quad \mathbf{F}_{\mathrm{sar}}^{\mathrm{pos}} = \operatorname{Pos}(\mathbf{F}_{\mathrm{sar}}), 
\end{gathered}
\end{equation}
where $\operatorname{Pos}(\cdot)$ denotes the positional encoding operation. We have found 
this design effectively enhances feature similarity under large-scale geometric variations while resolving ambiguities induced by significant deformations, as quantitatively demonstrated in Table~\ref{tb:weighting_factors} of the ablation study section.

\noindent\textbf{Cross-Attention-Only Interaction.} The positionally encoded features subsequently undergo cross-attention operations where queries originate from one modality while keys and values derive from the other, formulated as: 
\begin{equation}
\begin{gathered}
\mathbf{Q}_{\mathrm{x}} = \mathbf{F}_{\mathrm{x}}^{\mathrm{pos}} \cdot \mathrm{W}_{\mathrm{q}},\quad   
\mathbf{K}_{\mathrm{x}} = \mathbf{F}_{\mathrm{x}}^{\mathrm{pos}} \cdot \mathrm{W}_{\mathrm{k}},\quad 
\mathbf{V}_{\mathrm{x}} = \mathbf{F}_{\mathrm{x}}^{\mathrm{pos}} \cdot \mathrm{W}_{\mathrm{v}}, \\
\mathbf{F}_{\mathrm{opt\leftarrow sar}}^{'} = \operatorname{CrossAttn}(\mathbf{Q}_{\mathrm{sar}}, \mathbf{K}_{\mathrm{opt}}, \mathbf{V}_{\mathrm{opt}}), \\
\mathbf{F}_{\mathrm{sar\leftarrow opt}}^{'} = \operatorname{CrossAttn}(\mathbf{Q}_{\mathrm{opt}}, \mathbf{K}_{\mathrm{sar}}, \mathbf{V}_{\mathrm{sar}}),
\end{gathered}
\end{equation}
where \( \mathrm{W}_{\mathrm{q}} \), \( \mathrm{W}_{\mathrm{k}} \), and \( \mathrm{W}_{\mathrm{v}} \) denote learnable weight matrices, \( \mathrm{x} \in \{\mathrm{opt}, \mathrm{sar}\} \) specifies the modality type, and \( \mathbf{F}_{\mathrm{opt\leftarrow sar}} \)/\( \mathbf{F}_{\mathrm{sar\leftarrow opt}} \) represent the refined optical/SAR deep features after cross-modal interaction. This hierarchical process selectively aggregates knowledge from potential matching candidates in another image by measuring cross-view feature similarity, achieving selective inter-modal information aggregation.
This process generates modality-interacted independent features \( \mathbf{F}_{\mathrm{opt\leftarrow sar}}^{\prime} \) and \( \mathbf{F}_{\mathrm{sar\leftarrow opt}}^{\prime} \). To further aggregate cross-modal latent information, we recursively apply the cross-attention mechanism:
\begin{equation}
\begin{aligned}
\mathbf{F}_{\mathrm{opt\leftarrow sar}}^{\prime\prime} &= \mathrm{CrossAttn}(\mathbf{Q}_{\mathrm{sar}}^{\prime}, \mathbf{K}_{\mathrm{opt}}^{\prime}, \mathbf{V}_{\mathrm{opt}}^{\prime}) \\
\mathbf{F}_{\mathrm{sar\leftarrow opt}}^{\prime\prime} &= \mathrm{CrossAttn}(\mathbf{Q}_{\mathrm{opt}}^{\prime}, \mathbf{K}_{\mathrm{sar}}^{\prime}, \mathbf{V}_{\mathrm{sar}}^{\prime})
\end{aligned}
\end{equation}

 These doubly refined features \( \mathbf{F}_{\mathrm{opt\leftarrow sar}}^{\prime\prime} \) and \( \mathbf{F}_{\mathrm{sar\leftarrow opt}}^{\prime\prime} \) serve as inputs for subsequent optical flow prediction. Notably, to mitigate the computational complexity inherent in pairwise attention operations, we adopt a shifted local window attention strategy consistent with GMFlow~ \cite{xu2023gmflow}, where the number of windows is fixed at 4. The proposed two-stage cross-attention-only architecture demonstrates superior efficacy over conventional self- and cross-attention frameworks, as cross-modal information interaction plays a more pivotal role than single-modality feature depth in heterogeneous image registration tasks, which is validated by ablation experiments in Section~\ref{sec：Ablation}.

\subsection{LSR Module}
\label{method-sub-B}
Optical flow fields inherently possess multiple degrees of freedom and can naturally simulate non-rigid deformations. However, due to the unique imaging characteristics of SAR, strict pixel-wise alignment between SAR and optical images is fundamentally unattainable. Precise extraction of non-rigid transformations may amplify localized errors. For instance, building structures in optical imagery may exhibit layover distortion (top-bottom inversion) in SAR images. Precise local non-rigid registration in such areas risks introducing geometric misalignments, which may compromise overall registration accuracy. Consequently, constraining the mathematical model to an affine transformation, rather than pursuing non-rigid deformation, better captures the global registration relationship between optical and SAR images.
The affine transformation matrix $\Phi $, encompassing translation, scaling, and rotation, is mathematically expressed as:
\begin{equation}
\mathbf{\Phi}= \begin{bmatrix}
\mu _{1} & \mu _{2} & \mu _{3} \\
 \mu _{4}& \mu _{5} &  \mu _{6}\\
\end{bmatrix}=\begin{bmatrix}
S_{x}\ast cos\left ( \theta  \right ) & -S_{x}\ast sin\left ( \theta  \right ) & T_{x} \\
 S_{y}\ast sin\left ( \theta  \right )& S_{y}\ast cos\left ( \theta  \right ) &  T_{y}\\
\end{bmatrix},
\label{eq:matrix}
\end{equation}
where $S_{x}$, $S_{y}$, $T_{x}$, $T_{y}$ denote the scaling factors and translational offsets along the $x$ and $y$ axes, respectively; $\theta$ represents the rotation angle. 

Compared to the six degrees of freedom in affine transformations, the redundant degrees of freedom inherent to optical flow fields inevitably introduce additional registration errors. Mathematically, three corresponding points suffice to uniquely solve the six parameters of an affine transformation matrix. The mismatch filtering strategies widely employed in keypoint matching, such as RANSAC, operate through random sampling of triple-point subsets to iteratively estimate optimal models. However, this approach fails to leverage the inherent density and smoothness characteristics of optical flow fields.
To address this limitation, we innovatively propose the LSR network module, as illustrated in the yellow highlighted region of Fig.~\ref{Framework}. The LSR module adaptively regresses affine transformation parameters by exploiting dense correspondences rather than sparse subsets, which enhances robustness without requiring laborious parameter tuning procedures.

For any pixel position $\left[x_{o}, y_{o}\right]$ in the image, the coordinate displacement vector under an affine transformation can be computed via the affine transformation matrix $\mathbf{\Phi}$. Specifically, each unique $\Phi $ uniquely defines a distinct optical flow field, mathematically expressed as:
\begin{equation}
\mathbf{F} \left ( x , y\right ) = 
\begin{bmatrix}
 { flow }_{x}\\{ flow }_{y}
\end{bmatrix}=\begin{bmatrix}
  \mu _{1}-1& \mu _{2} & \mu _{3}\\
  \mu _{4}& \mu _{5}-1 & \mu _{6}
\end{bmatrix}\begin{bmatrix}
 x_{o}\\y_{o}\\1
\end{bmatrix},
\label{eq:flow_matrix}
\end{equation}
where $ {flow }_{x}$, ${ flow }_{y}$ denote the vector magnitudes of the optical flow field along the x- and y-directions at coordinate $\left[x_{o}, y_{o}\right]$, respectively. When extending this transformation to all pixels in the image $I\in \mathbb{R}^{H\times W}$, the expression can be generalized as:
\begin{equation}
\mathbf{F}=\mathbf{X} \left (\boldsymbol{\Phi}-\mathbf{I}\right), 
\label{equ:F=XPhi}
\end{equation}
where $\mathbf{F}\in \mathbb{R}^{N\times 2} $ denotes the optical flow field, $\mathbf{X}\in \mathbb{R}^{N\times 3} $ represents the set of original pixel coordinates in the image,$\boldsymbol{\Phi}\in \mathbb{R}^{2\times 3} $ corresponds to the 6-parameter affine transformation matrix (comprising translation, scaling, and rotation), $\mathbf{I}\in \mathbb{R}^{2\times3 }$ is the identity matrix, and $N=H\times W$ denotes the total number of pixels. We formulate the parameter estimation as a least squares problem. By minimizing the residual sum of squares:
\begin{equation}
    \mathcal{L}(\mathbf{\Phi})=\|\mathbf{F}-\mathbf{X}(\boldsymbol{\Phi}-\mathbf{I})\|_{2}^{2},
\end{equation}
where the notation $\left \| \cdot  \right \| _{2} $ denotes the Euclidean norm.The optimal solution is obtained through solving the normal equations:
\begin{equation}
    \mathbf{X}^{T}\mathbf{F}=\mathbf{X}^{T} \mathbf{X} \left (\boldsymbol{\Phi}-\mathbf{I}\right),
\end{equation}
which yields the closed-form solution:
\begin{equation}
    \boldsymbol{\Phi} = \left ( \mathbf{X}^{T} \mathbf{X} \right ) ^{-1}\mathbf{X}^{T} \mathbf{\mathbf{F}} + \mathbf{I},
\end{equation}
where $\mathbf{\Phi}$ denotes the final network output, representing the affine transformation matrix optimized from the predicted optical flow.

\subsection{Training Configurations}
\label{method-sub-C}
Our training objective combines two complementary loss terms: an aggregated sequence loss and a geometric constraint loss. The aggregated sequence loss supervises the iterative flow refinement process of the GRU module by progressively weighting the flow estimates across iterations:  
\begin{equation}  
L_{\text{seq}} = \sum_{i=1}^{N} \omega^{N-i} \left\| f_{\text{os}}^i - f_{\text{os}}^{\text{gt}} \right\|_1,  
\end{equation}  
where $f_{os}^{i}$ denotes the estimated optical flow at i-th iteration, $f_{os}^{gt}$ represents the ground-truth flow from the optical image to the SAR image, and $\omega$ controls the temporal weighting decay. The exponentially decaying weights emphasize later iterations while maintaining gradient flow to earlier predictions, forming a coarse-to-fine optimization process.

The GRU-generated flow sequence $\left\{ f^{1},...,f ^{N}\right\}$ is geometrically regularized through proposed LSR module, producing corresponding affine transformations $\{\mathbf{\Phi}^1, ..., \mathbf{\Phi}^N\}$. We compute geometrically constrained flows $\left\{ f_{\text{lsr}}^{1},...,f_{\text{lsr}}^{N}\right\}$ via Eq.~\ref{eq:flow_matrix}, and evaluate the geometric constraint loss as:  
\begin{equation}  
L_{\text{geo}} = \sum_{i=1}^{N} \omega^{N-i} \left\| f_{\text{lsr}}^i - f^{\text{gt}} \right\|_1. 
\label{eq:geoloss}
\end{equation} 

The geometric constraint loss $L_{\text{geo}}$ imposes an affine transformation constraint on the predicted optical flow, which encourages to filter diverging mismatched points. The final training loss is a linear combination of the two loss terms:
\begin{equation}  
L_{\text{total}} = \lambda_{\text{seq}} \cdot L_{\text{seq}} + \lambda_{\text{geo}} \cdot L_{\text{geo}}.
\end{equation} 
In our experiments, the best overall registration accuracy and stability were achieved when $\lambda_{\text{seq}}=0.5$ and $\lambda_{\text{geo}}=0.5$. Setting $\lambda_{\text{seq}}=1$ and $\lambda_{\text{geo}}=0$ resulted in a slight degradation in performance, though the network remained relatively stable. Conversely, when $\lambda_{\text{seq}}=0$ and $\lambda_{\text{geo}}=1$, performance decreased significantly. Under the configuration with $\lambda_{\text{seq}}=0.5$ and $\lambda_{\text{geo}}=0.5$, the geometric loss helps refine the output structure and improves model performance by incorporating additional geometric constraints, building upon the foundation provided by the sequential loss. Additionally, the weighting coefficient $\omega$ was empirically set to 0.85, and the number of GRU iterations $N$ was set to 12 to achieve a smooth convergence with moderate computational costs.

 \begin{figure}
\centering
\includegraphics[width=1.0\linewidth, trim=160pt 100pt 240pt 155pt, clip]{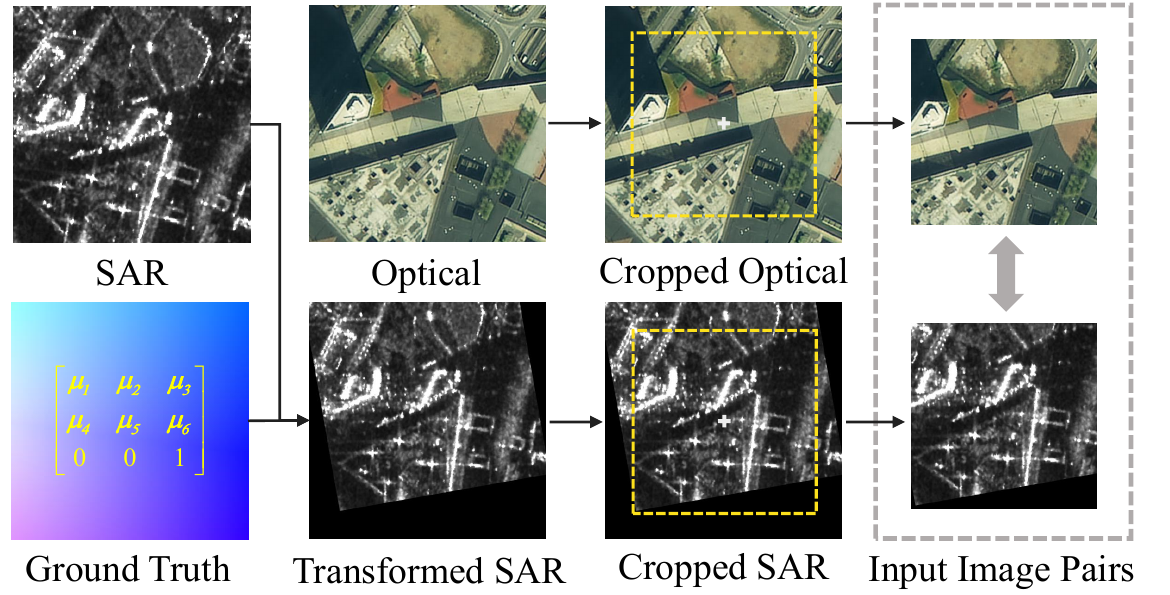}
\caption{Example of network input image pairs generation.}
\label{fig_detaset}
\end{figure}

\section{Experiments}
\label{Experiments}
\subsection{Benchmarks and Data Preparation}
\noindent\textbf{Benchmark Datasets:}
Our method was comprehensively validated on three publicly available datasets with different spatial resolutions: \textit{WHU-OPT-SAR dataset} \cite{li2022whu-opt-sardataset} (5-meter resolution), \textit{OS dataset} \cite{xiang2020osdataset} (1-meter resolution), and \textit{UBCv2 dataset} \cite{huang2023UBCV2dataset} (0.5-meter resolution). The WHU-OPT-SAR dataset contains 100 large-scale optical-SAR image pairs with different terrains, which are further segmented into 512×512 sub-images for the sake of efficiency: training (5,600 pairs), validation (700 pairs), and testing (700 pairs). The OS dataset spans multiple geographic regions and contains 2,673 aligned optical-SAR images, divided into splits of training (2,011 pairs), validation (238 pairs), and testing (424 pairs). The UBCv2 dataset, initially utilized for building detection and classification, contains 7,170 pairs of high-resolution optical and SAR satellite images. The high-resolution nature of UBCv2 data amplifies the texture differences of optical-SAR images and also results in much smaller field of view (FOV), serving as a representative benchmark for latest imaging resource. However, we notice that there are incomplete, cloud-occluded, or textureless image pairs within the UBCv2 dataset, which are unsuitable for registration tasks. Thus, we remove such data by a pre-screening step, and finally reach a split of training (3,517 pairs), validation (1,437 pairs), and test (1,447 pairs).

\noindent\textbf{Generation of Optical-SAR Image Pairs:}
The data preparation pipeline for optical-SAR image pairs used in network training is illustrated in Fig.~\ref{fig_detaset}. For each precisely registered image pair, we apply random affine transformations to the SAR image within specified parameter ranges to generate transformed SAR images, and compute its corresponding optical flow field as ground-truth supervision. Our experimental setup employs the following transformation bounds: The translation parameters were limited within the range of [-30, 30] with a precision of 1 pixel, the scaling parameter was limited within the range of [0.8, 1.2] with a precision of 0.05, and the rotation parameters were limited within the range of [-20°, 20°] with a precision of 1°. We believe such a transformation range presents significant technical challenges to optical-SAR registration, as most leading approaches are limited to translation-only or small-scale rotational/scaling transformations. 

Notably, our experiments reveal a positive correlation between the richness of shared structural information in image pairs and the registration accuracy. Although the original full-size input enriches contextual information, it substantially increases the computational load for flow predictions.
To further preserve shared image content and avoid interference from invalid black-border artifacts during applied transformations, we center-crop the 512×512 pixel input to a size of 400×400, as illustrated in Fig.~\ref{fig_detaset}.

\subsection{Metrics and Experimental Settings}
\textit{1) Evaluation Metrics:} To comprehensively assess registration performance, we employ three popular metrics and additionally propose a novel metric to evaluate the overall registration accuracy across multiple error tolerance levels:

\noindent\textbf{Average Endpoint Error (AEPE)} computes the mean Endpoint Error (EPE) across all image pairs in the test set, where for each image pair the EPE computes averaged pixel-wise Euclidean distance $l_{2}$ between predicted keypoints and their ground-truth correspondences:

\begin{table*}
\centering
\setlength{\fboxsep}{1pt}
\caption{Comparative results of different methods on three test sets of the \textbf{WHU-OPT-SAR dataset (5-meter resolution)} in `mean $\pm$ std' format. \textbf{Bold} indicates the best result, and \underline{underline} indicates the second best result.} 
\resizebox{\textwidth}{!}{
\begin{tabular}{llcccccccc}
\toprule 
\multirow{2}{*}{\textbf{Category}} & \multirow{2}{*}{\textbf{Method}} & \multicolumn{2}{c}{$\boldsymbol{\tau \leq 1}$\textbf{px}} & \multicolumn{2}{c}{$\boldsymbol{\tau \leq 2}$\textbf{px}} & \multicolumn{2}{c}{$\boldsymbol{\tau \leq 5}$\textbf{px}} & \multicolumn{2}{c}{\textbf{All}} \\
\cmidrule(lr){3-4} \cmidrule(lr){5-6} \cmidrule(lr){7-8} \cmidrule(lr){9-10}
 &  & CMR\text{@}$\tau$$\uparrow$ & AEPE\text{@}$\tau$$\downarrow$ & CMR\text{@}$\tau$$\uparrow$ & AEPE\text{@}$\tau$$\downarrow$& CMR\text{@}$\tau$$\uparrow$ & AEPE\text{@}$\tau$$\downarrow$ & AEPE$\downarrow$ & RMSE$\downarrow$ \\
\midrule
\multirow{3}{*}{\textbf{Sparse}} & RIFT2 \cite{li2019rift, li2023rift2} & 
  0.14{\footnotesize ($\pm$0.12)}\% & 
  \textbf{0.58{\footnotesize ($\pm$0.41)}} &  
  1.53{\footnotesize ($\pm$0.24)}\% & 
  1.49{\footnotesize ($\pm$0.05)} &  
  8.52{\footnotesize ($\pm$0.44)}\% & 
  3.04{\footnotesize ($\pm$0.05)} &  
  192.58{\footnotesize ($\pm$1.93)} & 
  15524.13{\footnotesize ($\pm$374.22)} \\

 & LNIFT \cite{li2022lnift} & 
  0.14{\footnotesize ($\pm$0.00)}\% & 
  0.84{\footnotesize ($\pm$0.07)} &  
  1.33{\footnotesize ($\pm$0.47)}\% & 
  1.45{\footnotesize ($\pm$0.06)} &  
  10.00{\footnotesize ($\pm$0.35)}\% & 
  3.19{\footnotesize ($\pm$0.10)} &  
  159.10{\footnotesize ($\pm$3.76)} & 
  11479.10{\footnotesize ($\pm$95.99)} \\

 & XoFTR \cite{tuzcuouglu2024xoftr} & 
  5.27{\footnotesize ($\pm$1.01)}\% & 
  0.79{\footnotesize ($\pm$0.01)} &  
  24.38{\footnotesize ($\pm$0.96)}\% & 
  1.30{\footnotesize ($\pm$0.02)} &  
  30.91{\footnotesize ($\pm$1.11)}\% & 
  \underline{1.53{\footnotesize ($\pm$0.03)}} & 
  53.00{\footnotesize ($\pm$1.58)} & 
  2874.67{\footnotesize ($\pm$49.59)} \\
\midrule
\multirow{6}{*}{\textbf{Dense}} & FlowFormer \cite{huang2022flowformer} & 
  14.29{\footnotesize ($\pm$1.52)}\% & 
  0.83{\footnotesize ($\pm$0.01)} & 
  64.81{\footnotesize ($\pm$0.99)}\% & 
  1.31{\footnotesize ($\pm$0.01)} &  
  91.81{\footnotesize ($\pm$0.41)}\% & 
  1.75{\footnotesize ($\pm$0.03)} &  
  2.98{\footnotesize ($\pm$0.20)} & 
  28.35{\footnotesize ($\pm$7.88)} \\

 & FlowFormer++ \cite{shi2023flowformer++} & 
  16.86{\footnotesize ($\pm$0.71)}\% & 
  0.81{\footnotesize ($\pm$0.01)} & 
  \underline{66.33{\footnotesize ($\pm$0.75)}\%} & 
  1.28{\footnotesize ($\pm$0.02)} & 
  90.00{\footnotesize ($\pm$0.84)}\% & 
  1.67{\footnotesize ($\pm$0.02)} &  
  3.22{\footnotesize ($\pm$0.18)} & 
  37.00{\footnotesize ($\pm$5.50)} \\

 & GMFlow \cite{xu2023gmflow} & 
  0.53{\footnotesize ($\pm$0.24)}\% & 
  0.93{\footnotesize ($\pm$0.06)} & 
  13.24{\footnotesize ($\pm$0.94)}\% & 
  1.59{\footnotesize ($\pm$0.01)} & 
  66.48{\footnotesize ($\pm$1.48)}\% & 
  3.01{\footnotesize ($\pm$0.00)} &  
  6.40{\footnotesize ($\pm$0.54)} & 
  249.73{\footnotesize ($\pm$120.83)} \\

 & RAFT \cite{teed2020raft} & 
  13.91{\footnotesize ($\pm$0.44)}\% & 
  0.81{\footnotesize ($\pm$0.01)} & 
  61.86{\footnotesize ($\pm$0.20)}\% & 
  1.31{\footnotesize ($\pm$0.01)} & 
  \underline{98.33{\footnotesize ($\pm$0.18)}\%} & 
  1.88{\footnotesize ($\pm$0.02)} & 
  \underline{2.04{\footnotesize ($\pm$0.07)}} & 
  \underline{4.25{\footnotesize ($\pm$3.22)}} \\

 & OS$^3$Flow \cite{sun2024os3flow} & 
  \underline{21.14{\footnotesize ($\pm$0.12)}\%} & 
  0.72{\footnotesize ($\pm$0.01)} &  
  57.19{\footnotesize ($\pm$1.98)}\% & 
  \underline{1.19{\footnotesize ($\pm$0.00)}} & 
  93.38{\footnotesize ($\pm$0.37)}\% & 
  1.89{\footnotesize ($\pm$0.04)} & 
  2.35{\footnotesize ($\pm$0.12)} & 
  6.65{\footnotesize ($\pm$3.09)} \\

 & \textbf{Ours} & 
  \textbf{72.05{\footnotesize ($\pm$1.06)}\%} & 
  \underline{0.60{\footnotesize ($\pm$0.01)}} & 
  \textbf{96.86{\footnotesize ($\pm$0.65)}\%} & 
  \textbf{0.78{\footnotesize ($\pm$0.01)}} & 
  \textbf{99.57{\footnotesize ($\pm$0.11)}\%} & 
  \textbf{0.83{\footnotesize ($\pm$0.01)}} & 
  \textbf{0.90{\footnotesize ($\pm$0.04)}} & 
  \textbf{0.62{\footnotesize ($\pm$0.15)}} \\
\bottomrule
\end{tabular}
}
\label{tb:WHU-OPT-SARdataset_compare}
\end{table*}

\begin{figure*}[!tb]
\centering
\includegraphics[width=1.0\linewidth, trim=130pt 40pt 20pt 40pt, clip]{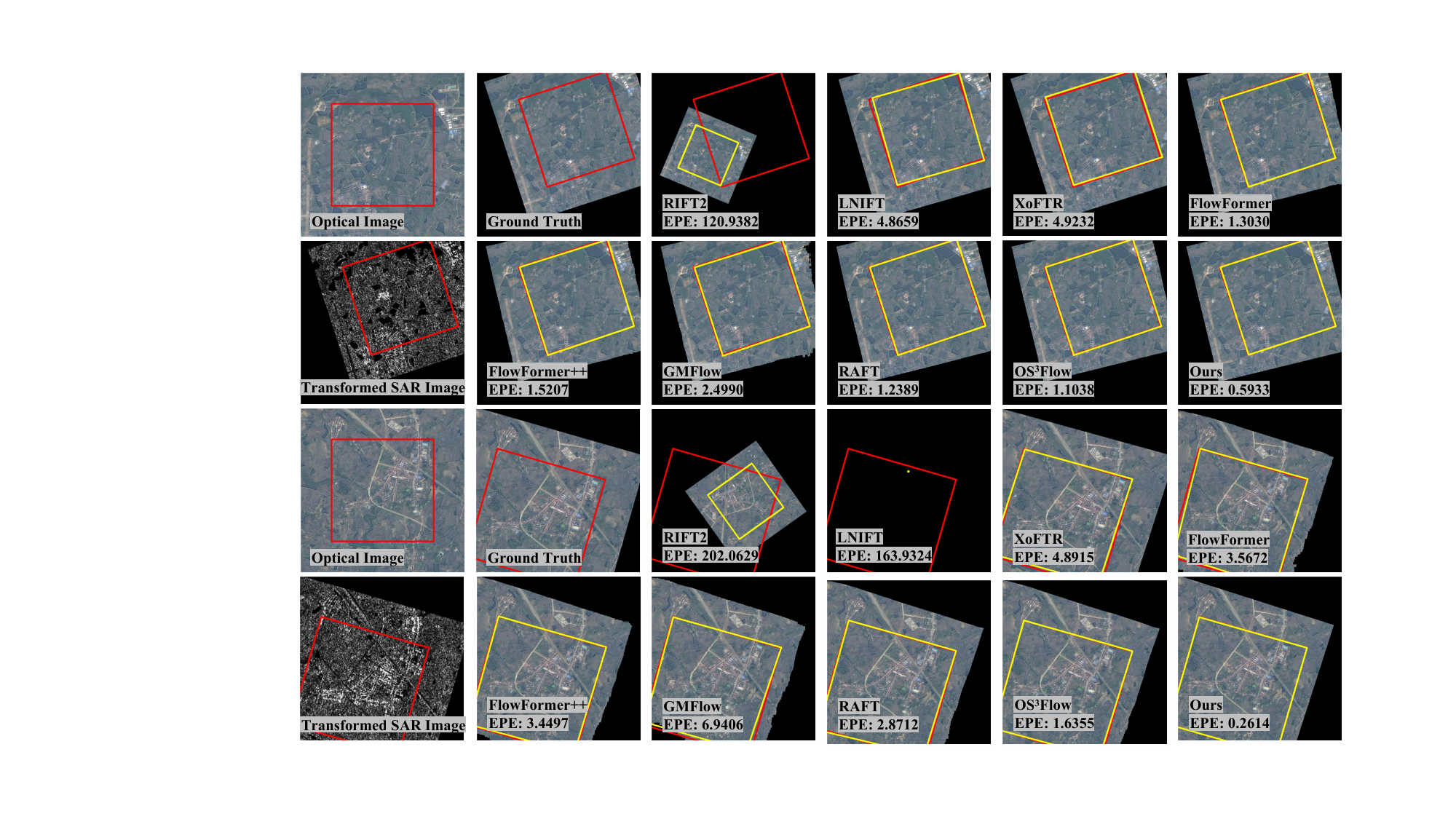}
\vspace{-16pt}
\caption{Registration results on the WHU-OPT-SAR dataset. The yellow line represents the ground truth registration result, and the red line represents the experimental registration result.}
\vspace{-15pt}
\label{fig:whu_results}
\end{figure*}

\begin{equation}
\begin{gathered}
\text{EPE}\left ( k \right )  = \frac{1}{M} \sum_{(x,y) \in I_{opt}^{k}}l_{2}(x,y),
\\
l_{2} =  \sqrt{ \left( f_{u}^{\text{pre}} - f_{u}^{\text{gt}} \right)^2 + \left( f_{v}^{\text{pre}} - f_{v}^{\text{gt}} \right)^2 },
\end{gathered}
\end{equation}
where $f_{u}^{\text{pre}},f_{v}^{\text{pre}}$ and $f_{u}^{\text{gt}},f_{v}^{\text{gt}}$ denote the predicted and ground-truth optical flow vectors at pixel position $\left (x,y  \right ) $, respectively. $M$ is the total number of valid pixels within the reference optical image $I_{opt}$. As $\text{EPE}\left ( k \right )$ is computed for the $k$-th image pair, we could further derive AEPE using all per-pair EPE value of the test data:
\begin{equation}
    \operatorname{AEPE}=\frac{1}{N} \sum_{k=1}^{N}  EPE(k),
\end{equation}
where $N$ is the total number of image pairs in test set.

\noindent\textbf{Root Mean Square Error (RMSE)} measures the global variance of displacement magnitude across all image pairs, thereby evaluating the dispersion of registration accuracy:
\begin{equation}
\text{RMSE} = \sqrt{ \frac{1}{N}\sum_{k=1}^{N}\left (\text{EPE}\left ( k \right ) - \overline{\text{EPE}}\right)^2 },
\end{equation}
where $\text{EPE}\left ( k \right )$ is the endpoint error of the $k$-th image pair described above.

\noindent\textbf{Correct Match Rate\text{@}$\tau$ (CMR\text{@}$\tau$)} quantifies the proportion of correctly matched image pairs under predefined precision thresholds:  
\begin{equation}
    \operatorname{CMR}\text{@}\tau=\frac{N_{\tau}}{N_{\text {total }}}\times 100\% ,
\end{equation}
where $\tau$ denotes the precision threshold. $N_{\tau}=N_{\left\{k \mid EPE(k)<\tau\right\}}$ represents the number of image pairs satisfying $\text{EPE}<\tau$ in the test set. Following the standard practice in classical registration benchmarks  \cite{li2019rift,teng2023omird}, we employ a multi-threshold strategy to assess registration performance at different precision levels: for coarse-level matching evaluation, we adopt thresholds of $\tau=3px$ and $\tau=5px$, while for fine-level accuracy assessment, we utilize more strict thresholds of $\tau=1px$ and $\tau=2px$.

\noindent\textbf{Average Endpoint Error\text{@}$\tau$} (AEPE\text{@}$\tau$) is newly proposed to evaluate the overall registration accuracy under varying tolerance thresholds, complementing the limitation of CMR\text{@}$\tau$ which focuses solely on the image quantity within threshold $\tau$ instead of their absolute registration precision. Thus the AEPE\text{@}$\tau$ is defined as:
\begin{equation}
    \operatorname{AEPE}\text{@}\tau=\frac{1}{N_{\tau}} \sum_{k \in \mathcal{M}_{\tau}} EPE(k) ,
\end{equation}
where $\mathcal{M}_{\tau}=\left\{k \mid EPE(k)<\tau\right\}$. This allows us to characterize how matching accuracy evolves with precision thresholds, revealing further performance details and resilience under varying CMR\text{@}$\tau$ value. 
It should be noted that since AEPE\text{@}$\tau$ is threshold-dependent and influenced by the number of image pairs meeting the condition, it should not be interpreted in isolation. A meaningful evaluation of the registration performance for the subset of correctly matched samples can only be achieved by jointly analyzing AEPE\text{@}$\tau$ with the CMR.

\begin{table*}
\centering
\setlength{\fboxsep}{1pt}
\caption{Comparative results of different methods on three test sets of the \textbf{OS dataset (1-meter resolution)} in `mean $\pm$ std' format. \textbf{Bold} indicates the best result, and \underline{underline} indicates the second best result.}
\resizebox{\textwidth}{!}{
\begin{tabular}{llcccccccc}
\toprule 
\multirow{2}{*}{\textbf{Category}} & \multirow{2}{*}{\textbf{Method}} & \multicolumn{2}{c}{$\boldsymbol{\tau \leq 1}$\textbf{px}} & \multicolumn{2}{c}{$\boldsymbol{\tau \leq 2}$\textbf{px}} & \multicolumn{2}{c}{$\boldsymbol{\tau \leq 5}$\textbf{px}} & \multicolumn{2}{c}{\textbf{All}} \\
\cmidrule(lr){3-4} \cmidrule(lr){5-6} \cmidrule(lr){7-8} \cmidrule(lr){9-10}
 &  & CMR\text{@}$\tau$$\uparrow$ & AEPE\text{@}$\tau$$\downarrow$ & CMR\text{@}$\tau$$\uparrow$ & AEPE\text{@}$\tau$$\downarrow$& CMR\text{@}$\tau$$\uparrow$ & AEPE\text{@}$\tau$$\downarrow$ & AEPE$\downarrow$ & RMSE$\downarrow$ \\
\midrule
\multirow{3}{*}{\textbf{Sparse}} & RIFT2 \cite{li2019rift,li2023rift2} & 
  0.55{\footnotesize ($\pm$0.29)}\% & 
  0.92{\footnotesize ($\pm$0.06)} &  
  4.40{\footnotesize ($\pm$0.40)}\% & 
  1.48{\footnotesize ($\pm$0.14)} & 
  28.69{\footnotesize ($\pm$1.22)}\% & 
  3.13{\footnotesize ($\pm$0.03)} &  
  72.24{\footnotesize ($\pm$8.13)} & 
  12540.89{\footnotesize ($\pm$517.13)} \\

 & LNIFT \cite{li2022lnift} & 
  0.32{\footnotesize ($\pm$0.29)}\% & 
  \textbf{0.49{\footnotesize ($\pm$0.38)}} & 
  3.85{\footnotesize ($\pm$1.28)}\% & 
  1.52{\footnotesize ($\pm$0.05)} &  
  32.55{\footnotesize ($\pm$0.33)}\% & 
  3.29{\footnotesize ($\pm$0.05)} & 
  95.79{\footnotesize ($\pm$0.84)} & 
  12777.92{\footnotesize ($\pm$282.86)} \\

 & XoFTR \cite{tuzcuouglu2024xoftr} & 
  4.83{\footnotesize ($\pm$1.02)}\% & 
  0.81{\footnotesize ($\pm$0.00)} & 
  27.16{\footnotesize ($\pm$0.59)}\% & 
  \underline{1.36{\footnotesize ($\pm$0.05)}} & 
  56.23{\footnotesize ($\pm$1.33)}\% & 
  \underline{2.22{\footnotesize ($\pm$0.06)}} & 
  25.00{\footnotesize ($\pm$1.66)} & 
  2173.35{\footnotesize ($\pm$316.47)} \\
\midrule
\multirow{6}{*}{\textbf{Dense}} & FlowFormer \cite{huang2022flowformer} & 
  0.86{\footnotesize ($\pm$0.11)}\% & 
  0.91{\footnotesize ($\pm$0.02)} & 
  21.39{\footnotesize ($\pm$1.54)}\% & 
  1.57{\footnotesize ($\pm$0.02)} &  
  60.90{\footnotesize ($\pm$0.78)}\% & 
  2.61{\footnotesize ($\pm$0.06)} & 
  6.87{\footnotesize ($\pm$0.21)} & 
  103.70{\footnotesize ($\pm$14.15)} \\

 & FlowFormer++ \cite{shi2023flowformer++} & 
  0.16{\footnotesize ($\pm$0.11)}\% & 
  \underline{0.64{\footnotesize ($\pm$0.45)}} & 
  17.37{\footnotesize ($\pm$1.11)}\% & 
  1.62{\footnotesize ($\pm$0.01)} & 
  62.26{\footnotesize ($\pm$0.77)}\% & 
  2.76{\footnotesize ($\pm$0.06)} & 
  7.76{\footnotesize ($\pm$0.26)} & 
  112.10{\footnotesize ($\pm$9.78)} \\

 & GMFlow \cite{xu2023gmflow} & 
  0.00{\footnotesize ($\pm$0.00)}\% & 
  $-$ & 
  4.40{\footnotesize ($\pm$0.87)}\% & 
  1.65{\footnotesize ($\pm$0.06)} & 
  49.77{\footnotesize ($\pm$1.02)}\% & 
  3.36{\footnotesize ($\pm$0.05)} & 
  6.17{\footnotesize ($\pm$0.22)} & 
  22.42{\footnotesize ($\pm$7.30)} \\

 & RAFT \cite{teed2020raft} & 
  2.20{\footnotesize ($\pm$0.22)}\% & 
  0.85{\footnotesize ($\pm$0.03)} & 
  32.86{\footnotesize ($\pm$1.24)}\% & 
  1.54{\footnotesize ($\pm$0.02)} & 
  \underline{90.49{\footnotesize ($\pm$0.11)}\%} & 
  2.51{\footnotesize ($\pm$0.02)} & 
  \underline{2.95{\footnotesize ($\pm$0.07)}} & 
  \underline{5.44{\footnotesize ($\pm$3.94)}} \\

 & OS$^3$Flow \cite{sun2024os3flow} & 
  \underline{5.15{\footnotesize ($\pm$1.20)}\%} & 
  0.72{\footnotesize ($\pm$0.01)} & 
  \underline{31.70{\footnotesize ($\pm$3.94)}\%} & 
  1.42{\footnotesize ($\pm$0.04)} & 
  86.29{\footnotesize ($\pm$1.42)}\% & 
  2.50{\footnotesize ($\pm$0.11)} & 
  3.17{\footnotesize ($\pm$0.18)} & 
  7.07{\footnotesize ($\pm$4.60)} \\

 & \textbf{Ours} & 
  \textbf{33.88{\footnotesize ($\pm$0.73)}\%} & 
  0.73{\footnotesize ($\pm$0.00)} & 
  \textbf{80.19{\footnotesize ($\pm$1.77)}\%} & 
  \textbf{1.13{\footnotesize ($\pm$0.00)}} & 
  \textbf{99.45{\footnotesize ($\pm$0.11)}\%} & 
  \textbf{1.46{\footnotesize ($\pm$0.02)}} & 
  \textbf{1.48{\footnotesize ($\pm$0.02)}} & 
  \textbf{0.75{\footnotesize ($\pm$0.02)}} \\
\bottomrule
\end{tabular}
}
\label{tb:OSdataset_compare}
\end{table*}

\begin{figure*}[!tb]
\centering
\includegraphics[width=1.0\linewidth, trim=130pt 40pt 20pt 40pt, clip]{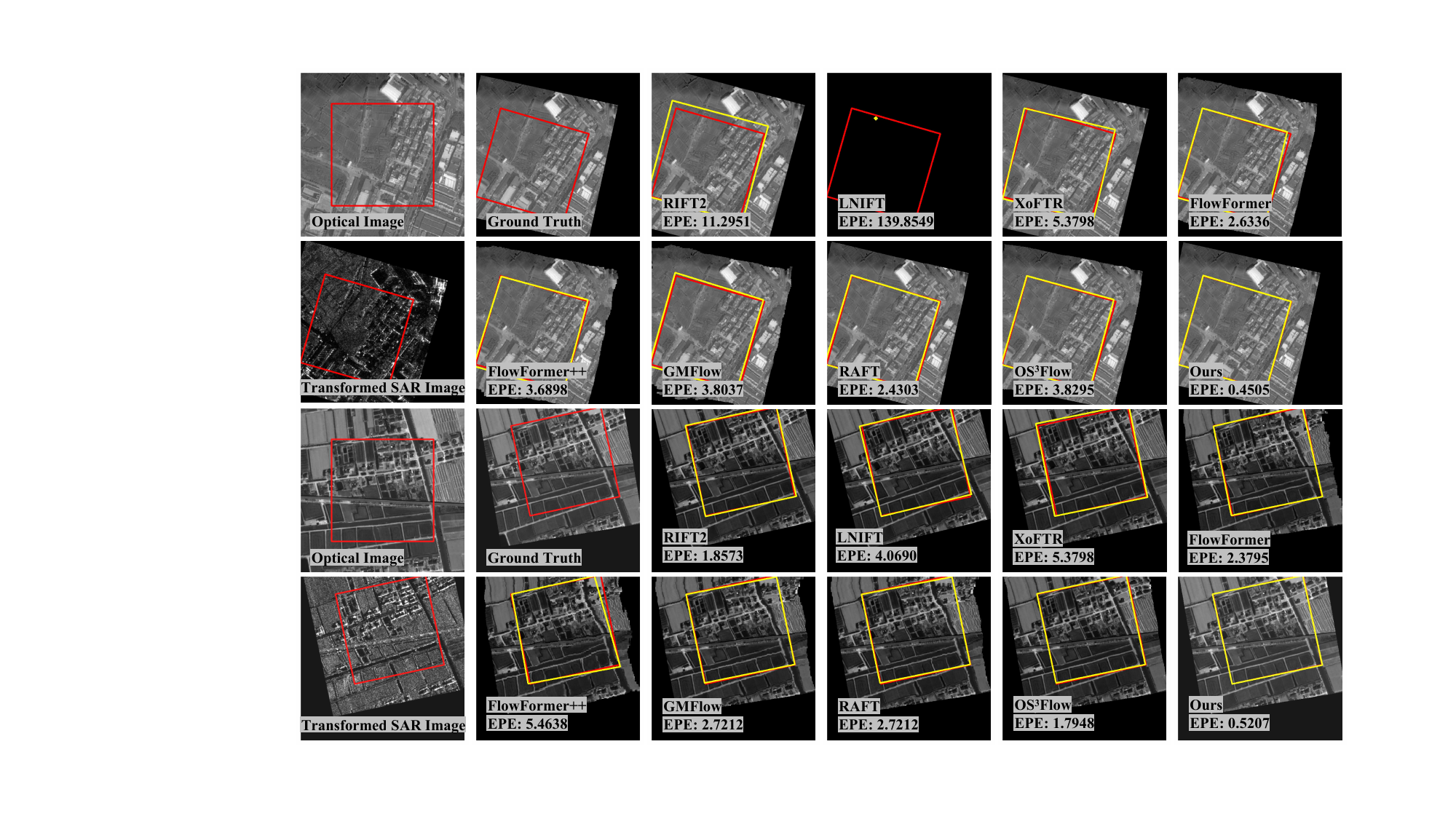}
\vspace{-16pt}
\caption{Registration results on the OS dataset. The yellow line represents the ground truth registration result, and the red line represents the experimental registration result.}
\vspace{-15pt}
\label{fig:os_results}
\end{figure*}

\textit{2) Experimental Settings:}
Our experiments utilized the AdamW optimizer for network training with an initial learning rate of 1.2e-5, a batch size of 12, and a maximum iteration of 120,000 steps. The GRU module underwent 12 iterations during training and 32 iterations during testing. When processing image pairs of size 512×512 pixels, the training process requires 16,672 MB of GPU memory. All experiments were implemented in PyTorch using a single NVIDIA GeForce RTX 4090 GPU and an Intel Core i9-14900k 24-core CPU.

All baseline methods are implemented by retraining their original pre-trained models using publicly released codebases with default training configurations. Notably, our cropping strategy, as illustrated in Fig.~\ref{fig_detaset}, was found to benefit both registration accuracy and robustness when applied to RAFT \cite{teed2020raft} and OS$^3$Flow \cite{sun2024os3flow}. Therefore, we incorporate this strategy into the implementations of RAFT and OS$^3$FLOW under identical experimental conditions to ensure fair comparisons.

Furthermore, to eliminate potential bias from the test dataset, we generate three independent test sets within predefined affine transformation ranges using different random seeds. All quantitative experimental results are reported as `mean $\pm$ standard deviation' across all three test sets.

\begin{table*}
\centering
\setlength{\fboxsep}{1pt}
\caption{Comparative results of different methods on three test sets of the \textbf{UBCv2 dataset (0.5-meter resolution)} in `mean $\pm$ std' format. \textbf{Bold} indicates the best result, and \underline{underline} indicates the second best result.}
\resizebox{\textwidth}{!}{
\begin{tabular}{llcccccccc}
\toprule 
\multirow{2}{*}{\textbf{Category}} & \multirow{2}{*}{\textbf{Method}} & \multicolumn{2}{c}{$\boldsymbol{\tau \leq 2}$\textbf{px}} & \multicolumn{2}{c}{$\boldsymbol{\tau \leq 3}$\textbf{px}} & \multicolumn{2}{c}{$\boldsymbol{\tau \leq 5}$\textbf{px}} & \multicolumn{2}{c}{\textbf{All}} \\
\cmidrule(lr){3-4} \cmidrule(lr){5-6} \cmidrule(lr){7-8} \cmidrule(lr){9-10}
 &  & CMR\text{@}$\tau$$\uparrow$ & AEPE\text{@}$\tau$$\downarrow$ & CMR\text{@}$\tau$$\uparrow$ & AEPE\text{@}$\tau$$\downarrow$& CMR\text{@}$\tau$$\uparrow$ & AEPE\text{@}$\tau$$\downarrow$ & AEPE$\downarrow$ & RMSE$\downarrow$ \\
\midrule
\multirow{3}{*}{\textbf{Sparse}} & RIFT2 \cite{li2019rift,li2023rift2} & 
  0.00{\footnotesize ($\pm$0.00)}\% & 
  $-$ & 
  0.00{\footnotesize ($\pm$0.00)}\% & 
  $-$ & 
  0.00{\footnotesize ($\pm$0.00)}\% & 
  $-$ & 
  236.79{\footnotesize ($\pm$1.80)} & 
  9253.72{\footnotesize ($\pm$110.04)} \\

 & LNIFT \cite{li2022lnift} & 
  0.00{\footnotesize ($\pm$0.00)}\% & 
  $-$ & 
  0.00{\footnotesize ($\pm$0.00)}\% & 
  $-$ & 
  0.14{\footnotesize ($\pm$0.00)}\% & 
  4.75{\footnotesize ($\pm$0.07)} & 
  224.28{\footnotesize ($\pm$0.21)} & 
  2018.86{\footnotesize ($\pm$41.38)} \\

 & XoFTR \cite{tuzcuouglu2024xoftr} & 
  0.00{\footnotesize ($\pm$0.00)}\% & 
  $-$ & 
  0.25{\footnotesize ($\pm$0.04)}\% & 
  2.89{\footnotesize ($\pm$0.01)} & 
  8.09{\footnotesize ($\pm$0.21)}\% & 
  4.01{\footnotesize ($\pm$0.11)} & 
  84.35{\footnotesize ($\pm$2.09)} & 
  3366.19{\footnotesize ($\pm$160.87)} \\
\midrule
 \multirow{6}{*}{\textbf{Dense}} & FlowFormer \cite{huang2022flowformer} & 
  0.00{\footnotesize ($\pm$0.00)}\% & 
  $-$ & 
  1.45{\footnotesize ($\pm$0.21)}\% & 
  2.59{\footnotesize ($\pm$0.05)} & 
  13.72{\footnotesize ($\pm$0.11)}\% & 
  4.05{\footnotesize ($\pm$0.02)} & 
  10.24{\footnotesize ($\pm$0.00)} & 
  40.48{\footnotesize ($\pm$1.11)} \\

 & FlowFormer++ \cite{shi2023flowformer++} & 
  0.00{\footnotesize ($\pm$0.00)}\% & 
  $-$ & 
  0.36{\footnotesize ($\pm$0.07)}\% & 
  2.56{\footnotesize ($\pm$0.10)} &  
  5.29{\footnotesize ($\pm$0.15)}\% & 
  4.10{\footnotesize ($\pm$0.00)} & 
  24.89{\footnotesize ($\pm$0.69)} & 
  261.65{\footnotesize ($\pm$15.06)} \\

 & GMFlow \cite{xu2023gmflow} & 
  0.00{\footnotesize ($\pm$0.00)}\% & 
  $-$ & 
  0.04{\footnotesize ($\pm$0.04)}\% & 
  \textbf{1.39{\footnotesize ($\pm$1.39)}} & 
  0.76{\footnotesize ($\pm$0.07)}\% & 
  4.30{\footnotesize ($\pm$0.01)} & 
  23.47{\footnotesize ($\pm$0.13)} & 
  160.79{\footnotesize ($\pm$5.21)} \\

 & RAFT \cite{teed2020raft} & 
  0.45{\footnotesize ($\pm$0.04)}\% & 
  1.76{\footnotesize ($\pm$0.14)} & 
  3.29{\footnotesize ($\pm$0.11)}\% & 
  2.51{\footnotesize ($\pm$0.01)} & 
  23.95{\footnotesize ($\pm$0.52)}\% & 
  3.94{\footnotesize ($\pm$0.02)} & 
  8.09{\footnotesize ($\pm$0.04)} & 
  \underline{23.39{\footnotesize ($\pm$1.17)}} \\

 & OS$^3$Flow \cite{sun2024os3flow} & 
  \underline{1.01{\footnotesize ($\pm$0.04)}\%} & 
  \underline{1.58{\footnotesize ($\pm$0.10)}} & 
  \underline{4.84{\footnotesize ($\pm$0.76)}\%} & 
  2.37{\footnotesize ($\pm$0.00)} & 
  \underline{26.75{\footnotesize ($\pm$1.18)}\%} & 
  \underline{3.77{\footnotesize ($\pm$0.04)}} & 
  \underline{7.99{\footnotesize ($\pm$0.08)}} & 
  23.81{\footnotesize ($\pm$1.95)} \\

 & \textbf{Ours} & 
  \textbf{14.89{\footnotesize ($\pm$0.38)}\%} & 
  \textbf{1.56{\footnotesize ($\pm$0.02)}} & 
  \textbf{37.08{\footnotesize ($\pm$0.04)}\%} & 
  \underline{2.13{\footnotesize ($\pm$0.00)}} & 
  \textbf{72.50{\footnotesize ($\pm$0.42)}\%} & 
  \textbf{2.97{\footnotesize ($\pm$0.00)}} & 
  \textbf{4.49{\footnotesize ($\pm$0.00)}} & 
  \textbf{11.97{\footnotesize ($\pm$0.31)}} \\
\bottomrule
\end{tabular}
}
\label{tb:UBCv2dataset_compare}
\end{table*}

\begin{figure*}[!tb]
\centering
\includegraphics[width=1.0\linewidth, trim=130pt 40pt 20pt 40pt, clip]{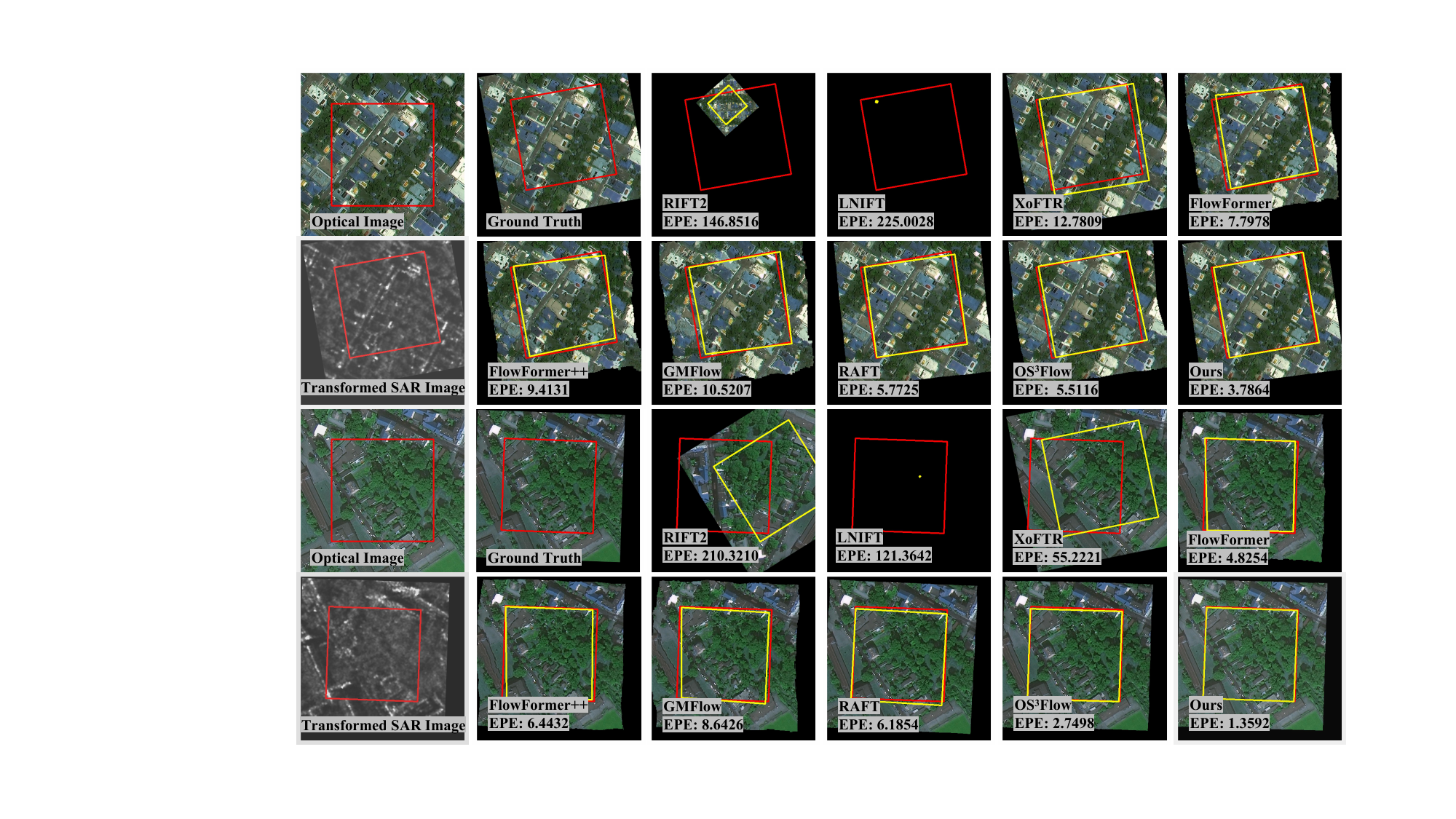}
\vspace{-16pt}
\caption{Registration results on the UBCv2 dataset. The yellow line represents the ground truth registration result, and the red line represents the experimental registration result.}
\label{fig:ubv2_results}
\end{figure*}

\subsection{Comparisons to Baseline Methods}
To validate the effectiveness of our method, we compare it against seven SOTA baseline methods, including three keypoint-based registration methods (RIFT2 \cite{li2019rift,li2023rift2}, LNIFT \cite{li2022lnift}, XoFTR \cite{tuzcuouglu2024xoftr}) and five dense-based registration methods (FlowFormer \cite{huang2022flowformer}, FlowFormer++ \cite{shi2023flowformer++}, GMFlow \cite{xu2023gmflow}, RAFT \cite{teed2020raft}, OS$^3$Flow \cite{sun2024os3flow}).   Qualitative and quantitative analyses were conducted on three distinct datasets to evaluate the generalization capabilities of these methods under varying spatial resolutions.

\textit{1) Results on the WHU-OPT-SAR dataset:} 
Qualitative comparisons on the WHU-OPT-SAR dataset (5m spatial resolution) are illustrated in Fig.~\ref{fig:whu_results}, where red bounding boxes denote ground-truth correspondences formed by connecting four reference points, and yellow boxes represent predictions from evaluated methods. Our method demonstrates exceptional alignment accuracy, with predicted yellow boxes nearly overlapping the red ground-truth boxes across diverse terrain scenarios.  

Quantitative results in Table~\ref{tb:WHU-OPT-SARdataset_compare} further validate our approach's superiority. Our method achieves \textbf{sub-pixel-level registration precision} with an overall AEPE of 0.90 pixels, surpassing the second-best method RAFT by 1.1 pixels. The advantages are particularly pronounced in high-precision matching metrics: at a threshold of $\tau \leq 1$ px, our method attains a CMR of 72.05\%, surpassing the suboptimal OS$^3$Flow by 50.91 percentage points. For $\tau \leq 2$ px, our CMR reaches 96.86\%, exceeding the suboptimal FlowFormer++ by 30.53 percentage points and covering nearly all test image pairs. Despite matching significantly more image pairs across thresholds, our method maintains nearly the lowest AEPE, demonstrating stable high-precision registration. Furthermore, our method achieves the lowest RMSE of 0.90, the only approach to fall below 1.0, underscoring its robustness and stability under low-resolution conditions. 

\textit{2) Results on the OS dataset:}
The OS dataset, with a spatial resolution of 1 m, presents significantly greater registration challenges compared to the WHU-OPT-SAR dataset. Under identical image dimensions, its effective receptive field captures $5^{2}$ times fewer cross-modal co-registered structural features, substantially increasing the difficulty of identifying shared correspondences. Qualitative results in Fig.~\ref{fig:os_results} demonstrate our method’s superior alignment accuracy, where predicted yellow bounding boxes closely align with ground-truth red boxes across diverse terrains, including urban areas and mountainous regions.  

\begin{table*}
\centering
\setlength{\fboxsep}{1pt}
\caption{Ablation study results of components in the feature extraction module on the OS dataset, including positional encoding, cross-attention, and self-attention.  \textbf{Bold} indicates the best result, and \underline{underline} indicates the second best result.}
\resizebox{\textwidth}{!}{
\begin{tabular}{cccccccccccccc}
\toprule 
\multicolumn{3}{c}{setup} & \multicolumn{2}{c}{All} & \multicolumn{2}{c}{$\tau \leq 1$px} & \multicolumn{2}{c}{$\tau \leq 2$px} & \multicolumn{2}{c}{$\tau \leq 3$px} & \multicolumn{2}{c}{$\tau \leq 5$px} & Param\\ 
\cmidrule(lr){1-3}\cmidrule(lr){4-5}\cmidrule(lr){6-7}\cmidrule(lr){8-9}\cmidrule(lr){10-11}\cmidrule(lr){12-13}
PE & SA & CA & AEPE$\downarrow$ & RMSE$\downarrow$ & CMR\text{@}$\tau$$\uparrow$ & AEPE\text{@}$\tau$$\downarrow$ & CMR\text{@}$\tau$$\uparrow$ & AEPE\text{@}$\tau$$\downarrow$ & CMR\text{@}$\tau$$\uparrow$ & AEPE\text{@}$\tau$$\downarrow$ & CMR\text{@}$\tau$$\uparrow$ & AEPE\text{@}$\tau$$\downarrow$ & $\left ( M \right )$ \\
\midrule
 & & \checkmark\checkmark & 2.45 & \underline{2.12} & 8.73\% & \underline{0.80} & 47.64\% & \underline{1.37} & 74.06\% & \textbf{1.75} & 92.92\% & \textbf{2.16} & 8.99\\
 \midrule
\checkmark & & & 2.99 & 24.64 & 4.25\% & \textbf{0.78} & 39.15\% & 1.44 & 70.99\% & 1.88 & 91.51\% & 2.31 & 5.32 \\
\checkmark & & \checkmark & 2.57 & \underline{2.12} & 6.84\% & 0.83 & 42.22\% & 1.39 & 70.75\% & 1.80 & 92.45\% & 2.28 & 7.16 \\
\checkmark & & \checkmark\checkmark\checkmark & \underline{2.43} & 2.15 & 8.25\% & \underline{0.80} & \underline{47.88\%} & \textbf{1.36} & \textbf{75.71\%} & \underline{1.76} & \underline{93.87\%} & \underline{2.17} & 10.82 \\
 \midrule
\checkmark & \checkmark & & 2.64 & 2.13 & 4.01\% & 0.85 & 40.80\% & 1.48 & 68.40\% & 1.88 & 92.92\% & 2.37 & 7.16 \\
\checkmark & \checkmark\checkmark & & 2.58 & 2.37 & 7.08\% & 0.81 & 41.98\% & 1.40 & 71.93\% & 1.85 & 93.63\% & 2.30 & 8.99 \\
\checkmark & \checkmark & \checkmark\checkmark & 2.56 & 3.58 & \textbf{10.38\%} & \underline{0.80} & 45.28\% & 1.38 & 72.88\% & 1.78 & 93.40\% & 2.23 & 10.82 \\
\midrule
\ding{52}& & \ding{52}\ding{52} & \textbf{2.40} & \textbf{1.76} & \underline{10.14\%} & 0.81 & \textbf{48.82\%} & 1.38 & \underline{74.29\%} & \textbf{1.75} & \textbf{94.34\%} & 2.19 & 8.99 \\ 
\bottomrule
\end{tabular}
}
\label{tb:weighting_factors}
\end{table*}

\begin{table}
\centering
\caption{Ablation results of the LSR module across three datasets. `LSR' denotes the least squares regression integrated into network training, and `LS' indicates the least squares regression excluded from network training.\textbf{Bold} indicates the best result, and \underline{underline} indicates the second best result.}
\resizebox{0.48\textwidth}{!}{
\begin{tabular}{cccccc}
\toprule
\multirow{2}{*}{Datasets} & \multirow{2}{*}{setup} & \multirow{2}{*}{AEPE$\downarrow$} & \multirow{2}{*}{RMSE$\downarrow$} &\multicolumn{2}{c}{CMR\text{@}$\tau$$\uparrow$} \\
\cmidrule(lr){5-6}  
 &  &  &  & $\tau \leq 1$px & $\tau \leq 2$px \\
\midrule
\multirow{2}{*}{WHU-OPT-} & None & 1.50{\footnotesize ($\pm$0.06)} & 4.35{\footnotesize ($\pm$2.36)} & 34.24{\footnotesize ($\pm$0.94)}\% & 83.86{\footnotesize ($\pm$0.77)}\% \\ 
\multirow{2}{*}{SAR dataset} & w/ LS & \textbf{0.90{\footnotesize ($\pm$0.03)}} & \underline{1.10{\footnotesize ($\pm$0.68)}} & \underline{71.24{\footnotesize ($\pm$0.94)}\%} & \underline{96.57{\footnotesize ($\pm$0.31)}\%} \\ 
 & w/ LSR & \underline{0.90{\footnotesize ($\pm$0.04)}} & \textbf{0.62{\footnotesize ($\pm$0.15)}} & \textbf{72.05{\footnotesize ($\pm$1.06)}\%} & \textbf{96.86{\footnotesize ($\pm$0.65)}\%} \\ 
 \midrule
\multirow{3}{*}{OS dataset} & None & 2.45{\footnotesize ($\pm$0.03)} & 1.85{\footnotesize ($\pm$0.04)} & 5.19{\footnotesize ($\pm$0.66)}\% & 46.62{\footnotesize ($\pm$1.49)}\% \\ 
 & w/ LS & \underline{1.50{\footnotesize ($\pm$0.06)}} & \underline{1.45{\footnotesize ($\pm$1.07)}} & \underline{32.00{\footnotesize ($\pm$1.25)}\%} & \underline{79.25{\footnotesize ($\pm$2.22)}\%} \\ 
 & w/ LSR & \textbf{1.48{\footnotesize ($\pm$0.02)}} & \textbf{0.75{\footnotesize ($\pm$0.02)}} & \textbf{33.88{\footnotesize ($\pm$0.73)}\%} & \textbf{80.19{\footnotesize ($\pm$1.77)}\%} \\ 
 \midrule
\multirow{3}{*}{UBCv2 dataset} & None & 7.33{\footnotesize ($\pm$0.01)} & 30.52{\footnotesize ($\pm$0.94)} & 0.00{\footnotesize ($\pm$0.00)}\% & 0.76{\footnotesize ($\pm$0.14)}\% \\ 
 & w/ LS & \underline{4.67{\footnotesize ($\pm$0.16)}} & \textbf{10.72{\footnotesize ($\pm$0.37)}} & \underline{0.11{\footnotesize ($\pm$0.11)}\%} & \underline{12.02{\footnotesize ($\pm$0.11)}\%} \\ 
 & w/ LSR & \textbf{4.49{\footnotesize ($\pm$0.00)}} & \underline{11.97{\footnotesize ($\pm$0.31)}} & \textbf{0.73{\footnotesize ($\pm$0.18)}\%} & \textbf{14.89{\footnotesize ($\pm$0.38)}\%} \\ 
\bottomrule`
\end{tabular}
}
\label{tab:mismatch removal}
\end{table}

\begin{figure*}[!htbp]
\centering
\includegraphics[width=0.9\linewidth, trim=65pt 100pt 40pt 80pt, clip]{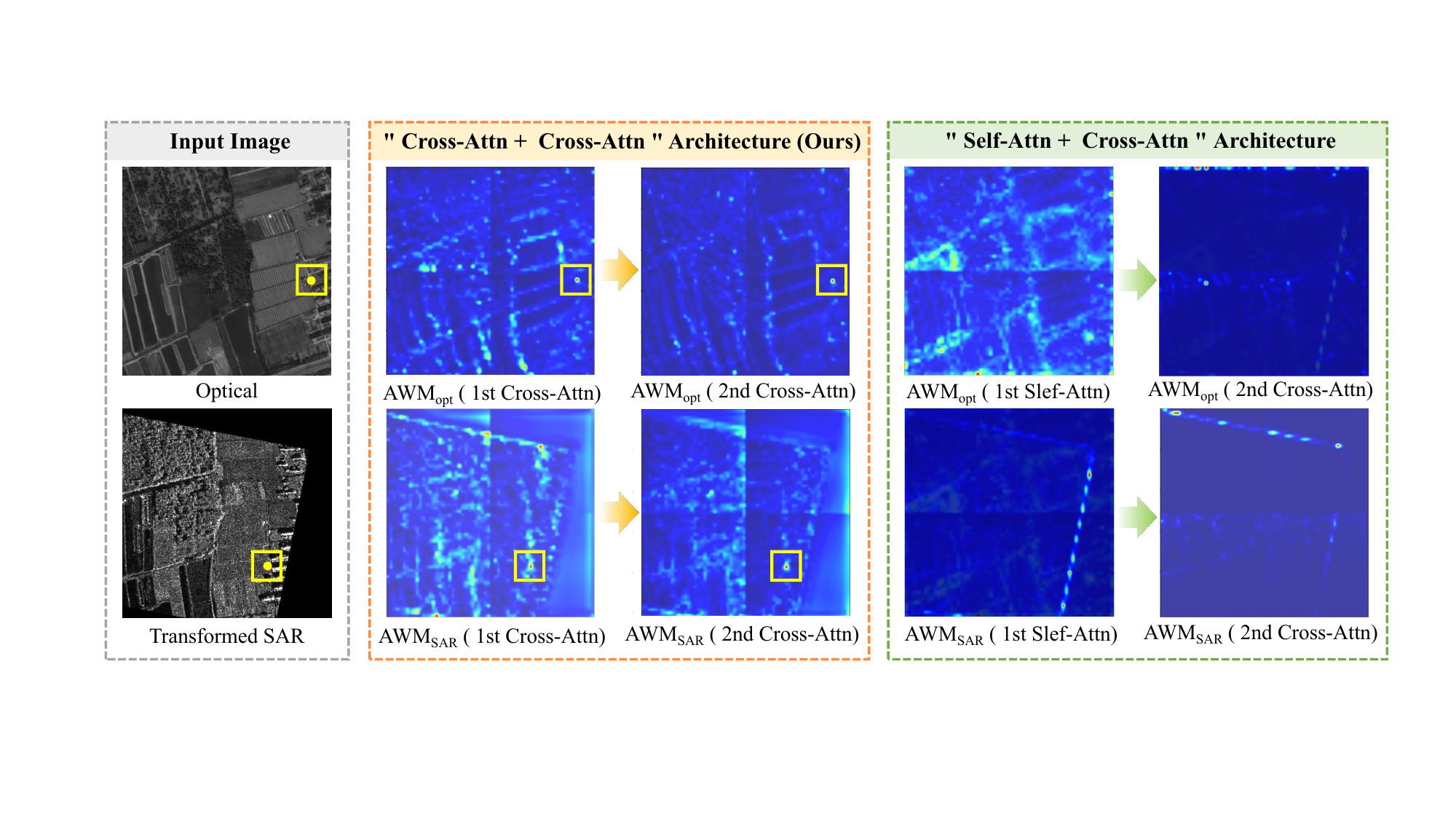}
\caption{Visualization of Attention Weight Matrices (AWM) from different attention architectures.}
\vspace{-15pt}
\label{fig:attention_feature_map}
\end{figure*}

\begin{figure*}[!tb]
\centering
\includegraphics[width=1.0\linewidth, trim=130pt 165pt 130pt 190pt, clip]{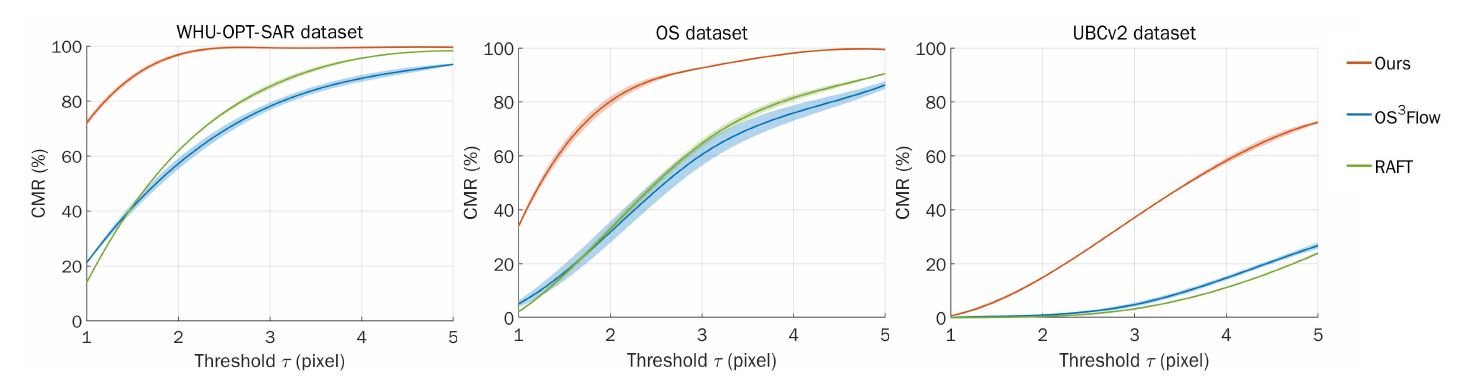}
\caption{CMR\text{@}$\tau$ metric with different thresholds on three benchmark datasets.}
\label{fig: CMR}
\end{figure*}

Table~\ref{tb:OSdataset_compare} presents the quantitative evaluation results of different methods on the OS dataset. Compared to the WHU-OPT-SAR dataset, the increased registration complexity of OS dataset leads to performance deterioration across all baseline methods. Nevertheless, our approach maintains significant superiority in both EPE and RMSE metrics. Specifically, Our method attains CMR of 33.88\% at $\tau \leq 1$ px, 80.19\% at $\tau \leq 2$ px, and 99.45\% at $\tau \leq 5$ px, surpassing the second-best methods by 28.73, 48.49, and 8.96 percentage points, respectively. It demonstrates that our method remains fully competent for coarse registration tasks on the moderately high spatial resolution OS dataset, while exhibiting substantially superior performance in high-precision correct registration rates compared to other SOTA approaches. Notably, our method attains an overall RMSE of approximately 1.5 pixels on OS dataset, further confirming its exceptional robustness in challenging registration scenarios.

\textit{3) Results on the UBCv2 dataset:}
The UBCv2 dataset, with an ultra-high spatial resolution of 0.5 m and fixed 512×512 image dimensions, contains very rare heterogeneous common structural features. Furthermore, there exist additional issues such as high image noise and cloud occlusion which severely degrade the quality of the input data. 
Compared with the WHU-OPT-SAR dataset and the OS dataset, the UBCv2 dataset exhibits substantially larger modality differences that pose new challenges for image registration. We therefore constrained the affine transformation parameters to narrower ranges: translation within [-15, 15] pixels (±1-pixel precision), scaling within [0.9, 1.1] (±0.05 precision), and rotation within [-10°, 10°] (±1° precision).

Qualitative results in Fig.~\ref{fig:ubv2_results} demonstrate that while alignment between predicted yellow boxes and ground-truth red boxes remains imperfect, our method exhibits marked improvements over competitors. Quantitative evaluations in Table~\ref{tb:UBCv2dataset_compare} reveal that traditional methods (RIFT, LNIFT) fail entirely on this dataset, while learning-based approaches suffer significant performance degradation. Nevertheless, our method achieves state-of-the-art results with CMR of 14.89\% at $\tau \leq 2$ px, 37.08\% at $\tau \leq 3$ px, and 72.50\% at $\tau \leq 5$ px, outperforming the second-best methods by 13.88, 32.24, and 45.75 percentage points, respectively.
The experimental results on the ultra-high-resolution UBCv2 dataset demonstrate that our method achieves coarse registration on the majority of image pairs and fine-grained registration on a subset of cases, despite the dataset’s extreme challenges. 

Fig.~\ref{fig: CMR} illustrates the trends of CMR under varying thresholds $\tau$ for different methods. The solid curves represent the mean CMR values of each method across three randomly transformed test sets, while the shaded regions around the curves indicate the variance in CMR observed across these sets.
It can be observed that our method significantly outperforms both RAFT and OS$^3$Flow in terms of both matching accuracy and stability. On the lower-resolution WHU-OPT-SAR dataset OS dataset, the advantage of our approach is most pronounced in high-precision matches with registration errors below 2 pixels. Moreover, it achieved nearly complete matching across almost all samples at thresholds of 3 and 4 pixels on the two datasets, respectively.
The results on the UBCv2 dataset further highlight the limitations of current cross-modal registration paradigms when applied to ultra-high-resolution scenarios. While our hybrid CNN-Transformer architecture and LSR module partially mitigate these challenges, the substantial performance gap emphasizes the need for novel methodologies to address severe modality discrepancies, sparse shared structural features,  and pervasive noise, which we leave as future research venue.

\begin{table*}
\centering
\setlength{\fboxsep}{1pt}
\caption{Comparison of Registration Efficiency Among Different Methods.}
\resizebox{\textwidth}{!}{
\begin{tabular}{ccccccccccc}
\toprule 
Method & RIFT2 & LNIFT & XoFTR & FlowFormer & FlowFormer++ & GMFlow & RAFT & OS$^3$Flow & Ours \\ 
\midrule
Param$\left ( M \right )$ & $-$ & $-$ & 10.86 & 16.08 & 15.88 & 4.68 & 5.25 & 10.52 & 8.99\\ 
FPS$\left (s / pair \right )$ & 12.9325 & 8.7453 & 0.1386 & 0.9271 & 1.5124 & 0.0983 & 0.1169 & 6.2885 & 0.1429 \\
GFLOPs & $-$ & $-$ & 293.77 & 146.96 & 71.07 & 63.45 & 207.01 & 414.03 & 273.62\\ 
\bottomrule
\end{tabular}
}
\label{tb:time_compare}
\end{table*}

\begin{figure*}[!tb]
\centering
\includegraphics[width=1.0\linewidth, trim=70pt 20pt 100pt 20pt, clip]{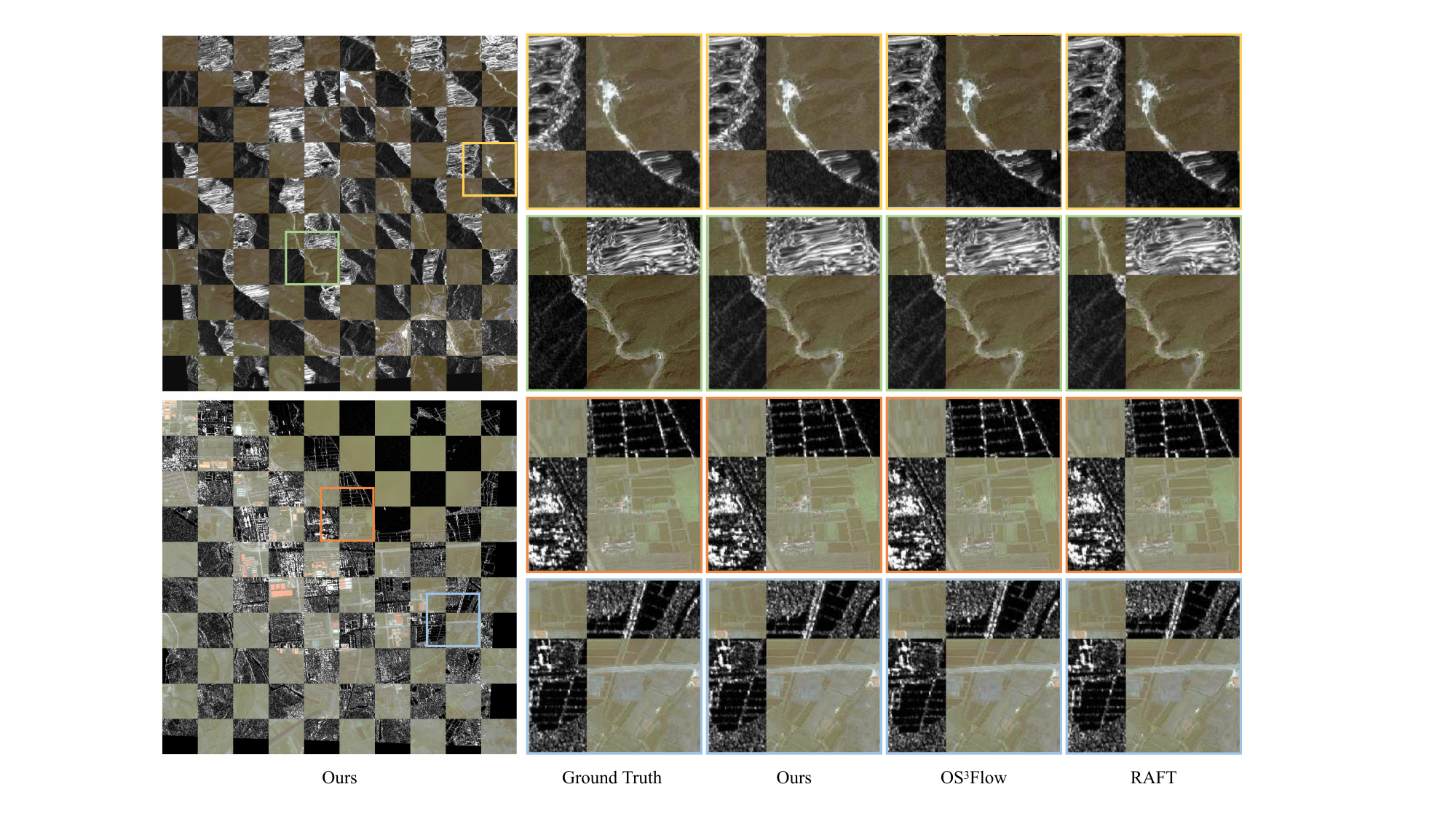}
\caption{Registration results of two large-scale image pairs (1600×1600 pixels) from the WHU-OPT-SAR dataset. Yellow, red, green, and blue boxes denote enlarged views of local regions covering distinct geographical features, including roads, harbors, and mountainous terrain.}
\vspace{-15pt}
\label{fig:chessboard}
\end{figure*}

\subsection{Ablation Study}
\label{sec：Ablation}
In this section, we conduct comprehensive ablation studies to analyze the contributions of key design choices in our method, including individual components within cross-modal feature extraction module and the LSR module.

\textit{1) Cross-Attention-Only Mechanism:}
To validate the effectiveness of our proposed cross-attention-only mechanism, we evaluated the impact of different components, including positional encoding (PE), cross-attention (CA), and self-attention (SA), on registration performance using the OS dataset, as summarized in Table~\ref{tb:weighting_factors}. Additionally, we compared the attention weight matrices produced by our proposed "dual-level cross-attention" architecture against those from the conventional "self-attention + cross-attention" structure, as illustrated in Fig.~\ref{fig:attention_feature_map}.

PE introduces spatial awareness into the registration process, enhancing robustness to geometric deformations. CA establishes inter-modal dependencies between optical and SAR features through selective information exchange. SA facilitates intra-modal context aggregation to refine domain-specific representations. As shown in the table, PE, CA, and SA contribute to improving registration performance to varying degrees. However, the "cross-attention only" configuration, comprising two cross-attention layers, is more suitable for optical-SAR registration than the conventional "SA + CA" design. This advantage can be visually interpreted from the attention weight matrices. In our "CA + CA" architecture (orange boxes in Fig.~\ref{fig:attention_feature_map}), the first cross-attention layer effectively filters and aligns cross-modal features, highlighting numerous potential correspondence regions. The second cross-attention layer further refines and fuses these aligned features at a deeper level, concentrating attention on semantically consistent key areas, as indicated by the yellow boxes. In contrast, in the traditional "SA + CA" architecture (green boxes in Fig.~\ref{fig:attention_feature_map}), the initial self-attention layer primarily enhances intra-image contextual relationships (e.g., structural details within a single building). However, due to the lack of cross-modal guidance at this stage, the resulting activations may not align well with the SAR modality. Given the significant domain gaps, including nonlinear radiometric differences, noise patterns, and structural discrepancies, self-attention maps from optical and SAR images often fail to achieve meaningful alignment. This misalignment can amplify modal differences rather than mitigate them, thereby impairing the subsequent cross-attention performance.

\textit{2) LSR Module:}
To validate the critical role of our proposed LSR module, we conducted systematic ablation experiments across three datasets. These experiments evaluated two configurations: the proposed LSR module integrated into network training, and the classical Least Squares (LS) post-processing excluded from network training. 

As detailed in Table~\ref{tb:weighting_factors}, both the LSR module and LS achieve significant performance improvements gains across all three benchmark datasets, compared to the original optical flow fields. This validates that imposing geometric constraints on divergent flow fields effectively filters error-prone correspondences. Notably, our LSR module delivers breakthrough improvements in sub-pixel precision metrics, increasing CMR ($\tau \leq 1$px) by more than 72\% on the WHU-OPT-SAR dataset.
Controlled comparative analyses further reveal that the LSR module achieves superior RMSE performance while maintaining robust registration performance on challenging image pairs, outperforming the LS baseline. This advantage originates from the geometric constraint loss (Eq.~\ref{eq:geoloss}), which enforces physically plausible affine transformations in optical flow predictions and suppresses outliers simultaneously. Experimental results demonstrate that the differentiable LSR module, guided by geometric priors during end-to-end network training, enhances both accuracy and robustness in cross-modal registration tasks. The synergistic optimization of geometric consistency and feature representation through our trainable architecture accounts for its consistent metric superiority across all datasets.

\subsection{Efficiency and Generalization Capability}
\textbf{Computational Efficiency.} Table~\ref{tb:time_compare} compares the performance of various methods in terms of parameter count, inference speed (FPS), and floating point operations (FLOPs). The FPS metric was computed by averaging the registration time per image pair over 700 test pairs. Our method achieves a processing speed of \textbf{0.1429 seconds per pair}, ranking second only to GMFlow, RAFT and XoFTR. FLOPs were measured using the thop library with input images resized to 512×512 pixels, yielding a computational complexity of 273.62 GFLOPs for our model. 
Notably, traditional methods RIFT/LNIFT exhibit significantly longer computation times compared to learning-based approaches, underscoring the computational superiority of deep learning paradigms. The results validate the practical viability of our framework for real-time cross-modal registration tasks. 

\textbf{Large-Scale Image Registration.} In practical registration tasks, the optical-SAR image pairs may have a large image size. To show the practicability of our method, we performed a direct evaluation on large-scale optical-SAR image pairs of size 1500×1500 using models trained exclusively on cropped 512×512 patches from the WHU-OPT-SAR dataset. The affine transformation ranges in these large-scale images remain consistent with those in the training images. Without loss of generality, we adopt a size of 1500x1500 as it is the largest dimension fitting our single GPU device. As visualized in the checkerboard comparison in Fig.~\ref{fig:chessboard}, our method achieves superior alignment accuracy in extended geographical features without any architecture modification or fine-tuning.  
More specifically, our method achieves exciting qualitative alignment accuracy in geographical patterns such as roads, coastal boundaries, and mountainous terrains, while competing methods exhibit obvious misalignment or blurring artifacts. Quantitatively, the registration achieved an EPE of 1.8 pixels and 1.4 pixels on two pairs of large-scale images, respectively, which validates the effectiveness of our framework in large-scale image registration tasks.

\section{Conclusion}
\label{Conclusion}
In this paper, we present GDROS, a geometry-constrained end-to-end framework for optical-SAR image registration. The proposed framework employs a hybrid CNN-Transformer architecture to extract cross-modality interactive deep features, followed by a novel LSR module that geometrically rectifies predicted optical flow fields while filtering out mismatched correspondences, ultimately deriving affine transformation parameters between image pairs. Extensive experiments on the WHU-OPT-SAR dataset, the OS dataset, and the UBCv2 dataset demonstrate GDROS's exceptional capability to address registration challenges under significant geometric variations, achieving high accuracy and robust performance across different resolution scenarios. While the framework shows promising scalability to large-scale images, its current limitations in high-precision registration for ultra-high-resolution data with sparse shared features highlight critical future research directions.

\section*{Acknowledgments}
Research presented here has been partly supported by the National Natural Science Foundation of China under Grant Number 62201603, the Hunan Provincial Natural Science Foundation of China under Grant Number 2025JJ40060, the China Postdoctoral Science Foundation under Grant Number 2023TQ0088, the Postdoctoral Fellowship Program of CPSF under Grant Number GZC20233539, and the Research Program of NUDT under Grant Number ZK22-04.

\begin{refcontext}[sorting=none]
\printbibliography
\end{refcontext}

\end{document}